\documentclass[a4paper,12pt]{article}
\pdfoutput=1 

\usepackage{jheppub} 

\usepackage[T1]{fontenc} 

\usepackage[percent]{overpic}
\usepackage{wrapfig}
\usepackage{tabu}
\usepackage{diagbox}

\usepackage{graphicx}
\usepackage{tikz}
\usepackage{import}
\usepackage{accents}
\usepackage{mathrsfs,amsmath,amssymb,slashed}
\usepackage{multirow,multicol}

\definecolor{darkgreen}{rgb}{0,0.3,0}
\definecolor{darkblue}{rgb}{0,0,0.3}
\definecolor{darkred}{rgb}{0.7,0,0}

\newcommand{\beqs}{\begin{equation*}}

\newcommand{\be}{\begin{equation}}
\newcommand{\ee}{\end{equation}}

\newcommand{\bome}{\boldsymbol{\omega}}
\newcommand{\hdid}{\hat{\text{d}}^\dagger}
\newcommand{\hdi}{\hat{\text{d}}}
\newcommand{\di}{\text{d}}
\newcommand{\did}{\text{d}^\dagger}
\newcommand{\beps}{\boldsymbol{\epsilon}}

\newcommand{\nn}{\nonumber}

\newcommand{\eqs}[1]{\begin{equation} #1 \end{equation}}

\newcommand{\eq}[2]{\begin{equation} #1 \label{#2} \end{equation}}

\newcommand{\F}{\mathcal{F}}
\newcommand{\A}{\mathcal{A}}
\newcommand{\bA}{\mathbf{A}}

\newcommand{\eps}{\varepsilon}

\newcommand{\de}{\delta}
\newcommand{\Laa}{\Lambda}
\newcommand{\laa}{\lambda}

\newcommand{\ex}{\text{x}}
\newcommand{\ordr}[1]{\mathcal{O}(\rho^{#1})}

\newcommand{\x}{\vec{x}}
\newcommand{\D}{\mathcal{D}}
\newcommand{\M}{\mathcal{M}}
\newcommand{\B}{\mathcal{B}}

\newcommand{\I}{\mathcal{I}}
\newcommand{\Ci}{\mathcal{C}}

\newcommand{\mc}[1]{\mathcal{#1}}
\newcommand{\pp}{{\prime\prime}}
\newcommand{\p}{\prime}

\newcommand{\blaa}{\boldsymbol{\laa}}
\newcommand{\blist}{\begin{itemize}}
\newcommand{\bcA}{\boldsymbol{\mathcal{A}}}
\newcommand{\bcF}{\boldsymbol{\mathcal{F}}}
\newcommand{\bcJ}{\boldsymbol{\mathcal{J}}}

\newcommand{\elist}{\end{itemize}}

\providecommand{\href}[2]{#2}

\DeclareFontFamily{OT1}{rsfs}{}

\DeclareFontShape{OT1}{rsfs}{m}{n}{ <-7> rsfs5 <7-10> rsfs7 <10->rsfs10}{} 

\DeclareMathAlphabet{\mycal}{OT1}{rsfs}{m}{n}

\DeclareMathOperator{\extdm}{d}

\newcommand{\extd}{\extdm \!}

\title{\centerline{\boldmath Asymptotic Symmetries in \emph{p}-Form Theories}}

\author[]{Hamid  Afshar,}
\author[]{Erfan Esmaeili, }
\author[]{M. M. Sheikh-Jabbari}

\affiliation[]{\it School of Physics, Institute for Research in Fundamental
Sciences (IPM),\\ P.O.Box 19395-5531, Tehran, Iran}

\emailAdd{afshar@ipm.ir}
\emailAdd{erfanili@ipm.ir}
\emailAdd{jabbari@theory.ipm.ac.ir}

\abstract{We consider $(p+1)$-form gauge fields in flat ($2p+4$)-dimensions for which radiation and Coulomb solutions have the same asymptotic fall-off behavior. Imposing appropriate fall-off behavior on fields and adopting a Maxwell-type action, we construct the boundary term which renders the action principle well-defined in the Lorenz gauge. We then compute conserved surface charges and the corresponding asymptotic charge algebra associated with nontrivial gauge transformations. We show that for $p\geq 1$, there are three sets of conserved asymptotic charges associated with \emph{exact}, \emph{coexact} and \emph{zero-mode} parts of the corresponding $p$-form  gauge transformations on the asymptotic $S^{2p+2}$. The coexact and zero-mode charges are higher form extensions of the four dimensional electrodynamics $(p=0)$, and are commuting. Charges associated with exact gauge transformations have no counterparts in four dimensions and form infinite copies of Heisenberg algebras. We briefly discuss physical implications of these charges and their algebra.}


\preprint{IPM/P-2018/002}

\begin{document} 
\maketitle
\flushbottom

\section{Introduction}\label{se:intro}

Theories of $(p+1)$-forms $\A_{\mu_0\cdots\mu_p}$ in $d$ dimensions, which we will denote by $(d,p)$-form theories, are natural extensions of the usual Maxwell theory of electrodynamics ($p=0,\ d=4$). Since multiplets of supersymmetry algebra generically contain $p$-forms, form field theories are ubiquitous in supersymmetric theories in $d>4$ \cite{Cremmer:1979up,Salam:1989fm}.  Two of the most famous examples are $p=even$ ($p=odd$) forms in 10d type IIb (IIa) supergravity and the $(6,1)$-form field theory in 6d supersymmetric theories \cite{Salam:1989fm, Ortin:2015hya, Polchinski:1998rr}. Being a part of short multiplets in such supersymmetric theories,  form fields of $(d,p)$-form theories always come with a gauge symmetry, where the gauge parameter is a generic $p$-form. This is a direct generalization of the case of Maxwell theory where the gauge parameter is a scalar. This $p$-form gauge symmetry manifests itself in the action of the theory which is expressed in terms of a $(p+2)$-form  field strength.
$(d,p)$-form gauge theories via their gauge fixing procedure in Lagrangian or Hamiltonian description and also their Dirac-type quantization conditions and duality properties have been extensively studied and analyzed in \cite{Deser:1997mz,Henneaux:1997ha,Teitelboim:1985yc,Teitelboim:1985ya,Banados:1997qs,Henneaux:1986ht,Cremmer:1997ct,Cremmer:1998px,Bremer:1997qb, Henneaux:1985kr, Baulieu:1987pz, Brown:1986nw}. Interactions and possible gauge groups involving $(d,p)$-forms  have also been studied \cite{Henneaux:1999ma,Bekaert:2000qx}. Moreover, $(d,p)$-form theories received a renewed attention after introduction of D$_p$-branes in string theory \cite{Polchinski:1995mt} as sources carrying the charge of $(p+1)$-form.

In this work, we progress further in study of $(d,p)$-form theories, their gauge symmetries and conserved charges.
Gauge symmetry is generically viewed as a manifestation of redundancy of the description in terms of gauge fields which should be removed and dealt with through gauge fixing procedure, e.g. see \cite{Henneaux:1986ht, Henneaux:1992ig}. Any gauge fixing procedure, however, leaves a residual part. Recalling the seminal Noether's theorem, one may ask if conserved charges can be associated with these residual gauge symmetries and what the physical meaning and implications of such charges is. Especially, recalling the lore that physical observables are made from gauge invariant quantities, does such a charge analysis have any relevance to physical observables? 

A place to look for such a physical meaning and implication is the Ward identity or BRST symmetry in gauge-fixed gauge theories. In  asymptotically flat spacetimes, an alternative formulation of Weinberg's soft theorems \cite{Weinberg:1965nx} has been proposed as Ward identities of these residual gauge symmetries with  non-trivial charges \cite{ Strominger:2013lka,Strominger:2013jfa,Kapec:2014opa,He:2014cra}. Such an explicit derivation of soft theorems from conserved quantities has been made in the context of gauge theories, gravity, higher spin theories and anti-symmetric $2$-form theories \cite{Kapec:2014zla,Kapec:2015ena,Campiglia:2015qka,Seraj:2016jxi,Conde:2016csj,DiVecchia:2017gfi,Campoleoni:2017mbt,Campoleoni:2017qot}. The analysis on conserved quantities are usually useful when we are only interested in comparing the state of a system at early  and late times. The memory effect  \cite{Christodoulou:1991cr} fits well into this setup as it amounts to capturing the traces remaining from  passage of a gravity or electromagnetic wave on the existing matter long after the wave has passed, without the need to follow the detailed evolution of the system \cite{Strominger:2014pwa,Pasterski:2015tva,Pasterski:2015zua,Susskind:2015hpa,Pate:2017fgt}.

For analyzing the residual gauge symmetries and associated conserved charges, there are some different systematic frameworks and formulations. The simplest one is based on the usual Noether's theorem suitably extended to capture these symmetries e.g. see \cite{Banados:2016zim, Avery:2015rga, Fatibene:1994vc}. There are other approaches based on Hamiltonian formulation e.g. see \cite{Brown:1986nw, Brown:1986ed, Regge:1974zd} and covariant phase space method \cite{Lee:1990nz, Barnich:2001jy}. To tackle the question of conserved charges in the $(d,p)$-form theory, which is a linear theory, we employ an appropriately extended version of the Noether's method which we find more handy. 

    In our case, we will be focusing on a specific $(d,p)$-form theory in flat Minkowski spacetime ${\cal M}_d={\mathbb R}^{d-1,1}$. In order to compute the conserved charges associated with residual (gauge) symmetries, we  choose to fix the Lorenz gauge which preserves Lorentz symmetry, and we specify the fall-off asymptotic behavior of gauge fields and/or residual gauge parameters. 
    Here we will choose the de Sitter slicing of the flat spacetime which is used in similar questions in 4d Maxwell theory or Einstein-Hilbert theory e.g. see \cite{Ashtekar:1991vb,Campiglia:2015qka} (The other two commonly used choices are slicing by constant time surfaces \cite{Troessaert:2017jcm, Campiglia:2017mua} and the null slicing \cite{He:2014cra, Barnich:2016lyg}, see below for more discussions.) We fix the fall-off behavior such that we find finite and well-defined expressions for the conserved surface charges in $d=2p+4$ dimensions. This spacetime dimension, as we will discuss, is special in some different ways; the most relevant one to our work in this paper being as follows. It is known that radiation flux of a localized source in $d$ dimensions has radial fall-off $r^{-(d-2)}$, which means the  field yielding this radiation should fall off as $r^{-\frac{d-2}{2}}$, usually called radiation fall-off behavior. On the other hand, the fall-off behavior of fields generated by `localized' electric sources, the so-called Coulomb fall-off behavior, for a $(p+1)$-form is $r^{-(d-p-3)}$ \cite{Ortaggio:2014ipa}. It has been argued that we have memory effect when these two are equal, i.e. $(d-2)/2=d-p-3$, which happens in $d=2p+4$.\footnote{See \cite{Hollands:2016oma,Pate:2017fgt,Campoleoni:2017qot,Satishchandran:2017pek,Garfinkle:2017fre,Mao:2017wvx,Campiglia:2017xkp} for discussions on (gravitational) memory effect in $d>4$ dimensions.}

    The surface charges are integrals over the asymptotic spheres $S^{d-2}=S^{2p+2}$ and the integrand is a linear function of the residual gauge transformation parameter, which is a harmonic $p$-form on ${\cal M}_d$. 
    The part of these harmonic $p$-forms which contribute to the surface charges is given by $p$-forms on the asymptotic $S^{2p+2}$ sphere. These forms can be decomposed into exact and  coexact parts, leading to two distinct sets of asymptotic surface charges which we conveniently call  \emph{exact and coexact asymptotic charges}. We then compute the asymptotic charge algebra and show that the coexact charges commute among themselves and also with exact charges. The exact charges, however, do not commute with each other and form Heisenberg algebras.

This paper is organized as follows. After fixing the conventions and notations in subsection \ref{sse:notation}, in section \ref{se:action} we introduce the $(d,p)$-form theory and present basic analysis of the theory, including gauge fixing, fall-off behavior, zero-mode charges and the boundary term established to make the action principle well-defined. Section \ref{section-3} contains our main analysis and results where we derive the expression for asymptotic conserved charges associated with residual gauge symmetries. We discuss that there are three classes of \emph{zero-mode, exact and coexact} charges and that the exact  sector, unlike the other two satisfies a non-Abelian charge algebra. In section \ref{sec:charge-computation-examples}, we present explicit computation of charges and their algebra for the two $p=0$ (4d Maxwell theory) and $p=1$ (6d $2$-form theory) cases.  We summarize and discuss our results in section \ref{discussion-section}. In appendices, we have gathered some technical details of our computations as well as other subsidiary approaches. In appendix \ref{Appendix-forms}, we briefly review differential forms and their Hodge decomposition on sphere by discussing separation of exact and coexact parts of the gauge fields and gauge parameters, needed for explicit computation of charges. Appendix \ref{Hamiltonian} concerns the Hamiltonian analysis of the $(d,p)$-form theory. This section provides complementary analysis to the action and Lagrangian descriptions of section \ref{se:action}.  In appendix \ref{se:covariant} we present the charge analysis in the covariant phase space method, as a complement to our `extended Noether's theorem' charge computations presented  in the main text.

\subsection{Notation and conventions}\label{sse:notation}

\paragraph{Coordinate systems and covariant derivatives.} We formulate the $(d,p)$-form gauge theories on $d$-dimensional Minkowski spacetime $\mathcal{M}_d$ with Cartesian coordinates $x^\mu$ and metric  $\eta_{\mu\nu}=\text{diag}(-+\cdots +)$, where Greek indices run over $0,\cdots,d-1$. We will mainly work in de Sitter slicing of Minkowski spacetime using the hyperbolic coordinate system  where the coordinates on $(d-1)$-dimensional de Sitter spacetime are denoted by $x^a$, and small Latin indices $(a,b,c,\cdots)$  run over $0,\cdots,d-2$. The line element on ${\cal M}_d$  in these two coordinate systems  is given by
\begin{equation}
ds^2=\eta_{\mu\nu} dx^\mu dx^\nu= -dt^2+dr^2+r^2d\Omega_{d-2}^2=d\rho^2+\rho^2h_{ab}\di x^a \di x^b\,,
\end{equation}
with $h_{ab}$ being the  metric on the unit-radius $(d-1)$-dimensional de Sitter space $dS_{d-1}$
\begin{equation}
 h_{ab}\di x^a\di x^b=-d\tau^2+\cosh^2\tau\, d\Omega_{d-2}^2\,\label{metric}.
\end{equation}
We note that the above metric describes a global $dS$ space.
The map between coordinates is
\begin{equation}
\rho^2=x^\mu x_\mu=r^2-t^2\,,\qquad \tanh\tau =\frac{t}{r}\,,\qquad \hat{x}\to \hat{x}\,,
\end{equation}
or equivalently,
\begin{equation}
r=\rho\cosh\tau\,,\qquad t=\rho\sinh\tau\,,   \qquad r,\rho\geq 0\,,\;\; t,\tau\in\mathbb{R}\,, 
\end{equation}
where $\hat{x}$ denotes a specific direction in space, corresponding to a point on unit $(d-2)$-sphere $S^{d-2}$. We will denote the coordinates on $S^{d-2}$ in either of de Sitter slicing or Minkowski coordinates by $x^A$, with the uppercase Latin indices ranging over $1,\cdots,d-2$ and the metric on ${\cal M}_d$, unit radius $dS_{d-1}$ and unit radius $S^{d-2}$ respectively by $g_{\mu\nu}, h_{ab}$ and $\mathscr{G}_{AB}$. The respective covariant derivatives are denoted by  $\nabla, D$ and $\mathcal{D}$ whose indices are raised and lowered as
\eqs{
\nabla^\nu=g^{\mu\nu}\nabla_\mu\,,\qquad
D^a=h^{ab}D_b\,,\qquad
\D^A=\mathscr{G}^{AB}\D_B\,.
}
Moreover, for the integration measures we use,
\eq{\sqrt{-h}=\sqrt{-\det{h_{ab}}}\,,\qquad \sqrt{\mathscr{G}}\equiv\cosh^{d-2}\tau\ \sqrt{\det{\mathscr{G}_{AB}}}\,. }{volume-measures}

\begin{figure}[t]
 \centering
 \begin{overpic}[width=0.2\textwidth,tics=1]{flat-penrose0.pdf}
 \put (28,25) {\large$\mathscr{I}^-$}
  \put (28,75) {\large$\mathscr{I}^+$}
   \put (52,48) {\large$i^0$}
 \put (-8,0) {\large$\displaystyle i^-$}
 \put (-8,96) {\large$\displaystyle i^+$}
\end{overpic}
\hspace{2.4cm}
 \centering
\begin{overpic}[width=0.2\textwidth,tics=1]{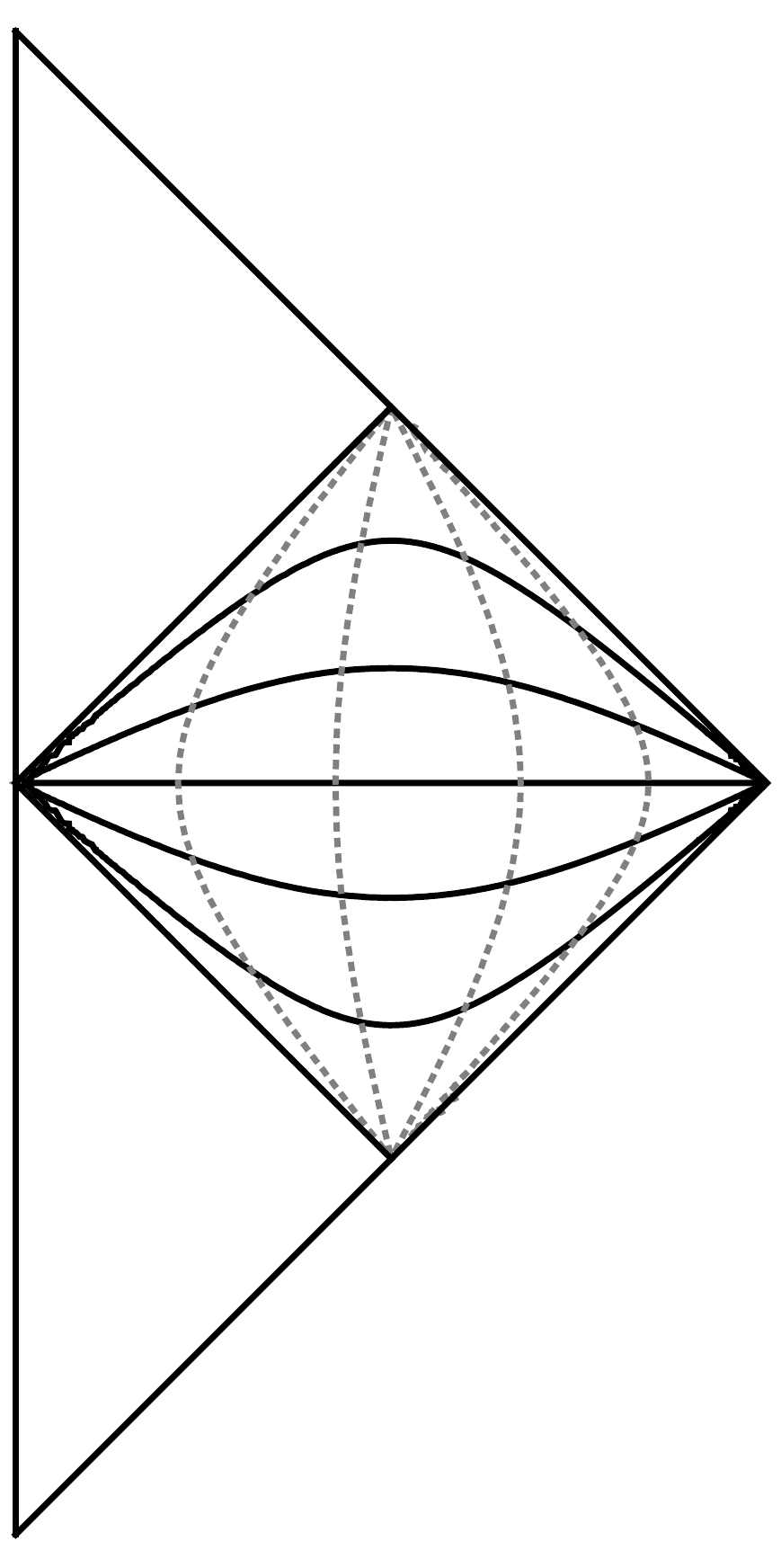}
 \put (28,25) {\large$\mathscr{I}^-$}
  \put (28,75) {\large$\mathscr{I}^+$}
   \put (52,48) {\large$i^0$}
 \put (-8,0) {\large$\displaystyle i^-$}
 \put (-8,96) {\large$\displaystyle i^+$}
 \end{overpic}
\caption{Penrose diagrams of Minkowski flat spacetime ${\cal M}_d$. (Left) The dashed and thick curves denote the constant Cartesian time $t$ and radial $r$ slices respectively. As we see all the constant $t$ curves meet the asymptotic spatial infinity $i_0$ with a zero slope. (Right) The patch covered by the de Sitter slicing. The solid lines are constant $\tau$ slices, while dotted lines are constant $\rho$ hyperboloids.  The de Sitter slicing does not cover future and past timelike infinities $i^\pm$. }
    \label{Mink-de-Stter-comparison}
\end{figure}
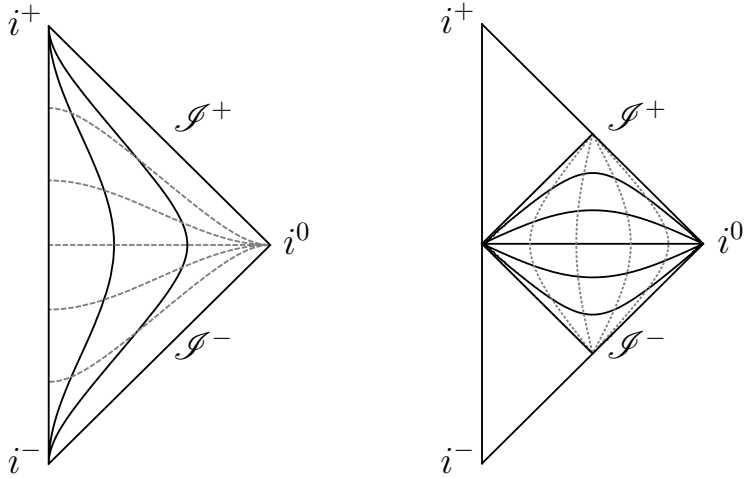

\begin{figure}[t]
 \centering
 \begin{tikzpicture}[thick,scale=0.5, every node/.style={scale=0.5}]




\draw[magenta,thick] (8,4.3)..controls(5.5,0)..(8,-4.3)node{};
\draw[magenta,thick] (-2,4.3)..controls(.5,0)..(-2,-4.3)node{};

\path[fill=magenta,opacity=0.1] (-2,4.3)..controls(.5,0)..(-2,-4.3)
 arc (180:360:5cm and 1cm)
 (8,-4.3)..controls(5.5,0)..(8,4.3)  arc (0:-180:5cm and 0.8cm)
 ;

\draw[draw=magenta] (3,4.3) ellipse (5cm and 0.8cm);
\draw[magenta] (-2,-4.3) arc (180:360:5cm and 1cm)node{};
\draw[magenta,very thick] (-0.13,0) arc (180:360:3.12 cm and .8cm);


\draw[dashed] (-2.2,5.2)--(8.2,-5.2);
\draw [dashed] (-2,-5.2)--(8.2,5.2);

\node [rotate=65] at (-1.2,-2) {\LARGE$\rho=\rho_0$};
\node at (3.7,0) {\LARGE$\mathcal{O}$};
\node  at (7,0) {\LARGE$\tau=0$};

\node [rotate=63] at (8.2,3.5) {\LARGE$\tau\to\infty$};
\node [rotate=-63] at (8.2,-3.5) {\LARGE$\tau\to-\infty$};

\end{tikzpicture}
    \caption{Embedding the de sitter space in global patch as $\rho$-constant slices of Minkowski pace with $\rho^2=x_\mu x^\mu$. The $\rho_0\to\infty$ region gives the $dS_{d-1}$ at the boundary of the Minkowski space. }
    \label{de-Sitter}
\end{figure}
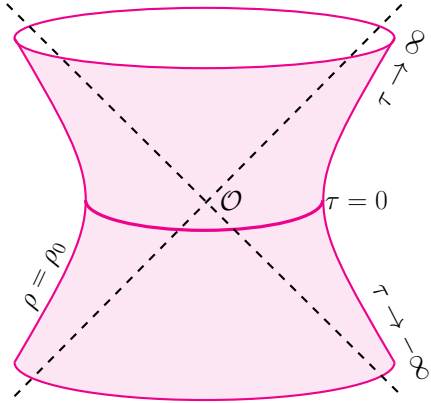

The de Sitter slicing coordinate system covers $r>|t|$ patch of Minkowski spacetime, foliated by codimension 1 hypersurfaces with de Sitter metric. What we mean by \emph{spatial boundary} is the de Sitter space at $\rho\to\infty$ \cite{Ashtekar:1991vb,Compere:2011db}.
The comparison between the Cartesian and the de Sitter slicing coordinates and their coverage on the Penrose diagram of the flat space has been
depicted in Fig. \ref{Mink-de-Stter-comparison}.  As is seen in the figure, the de Sitter slicing covers a part of the flat spacetime causally disconnected from the origin. Its `boundary' (large $\rho$ region) covers spatial infinity $i^0$ and up to half of past and future null infinities $\mathscr{{I}}^\pm$ depending on how the limit to asymptotic region is taken.
The de Sitter slicing makes manifest the $SO(d-1,1)$ Lorentz invariance of the Minkowski spacetime ${\cal M}_d$; at any constant $\rho$ slice we have a $d-1$ dimensional de Sitter spacetime  depicted in Fig. \ref{de-Sitter} whose isometry group is $SO(d-1,1)$. The reason is that $ \rho$ is a Lorentz invariant quantity and it is the proper distance of points from the origin of Minkowski space. The timelike coordinate $\tau$ on de Sitter slices determines the rate $t/r$ by which radial spacelike curves emanate from the origin.  The $\tau=0$ slice corresponds to $t=0$ region of the Minkowski and $\tau\to\pm\infty$  include parts of asymptotic null infinities $\mathscr{I}^\pm$.  
For later use we note that rigid scaling acting as $x^\mu\to\lambda x^\mu$ in the Minkowski coordinates, appears as $\rho\to\lambda\rho$ in the de Sitter slicing  with the other coordinates ($\tau, \hat x$) remaining intact (\emph{cf.} footnote \ref{scaling footnote}). 

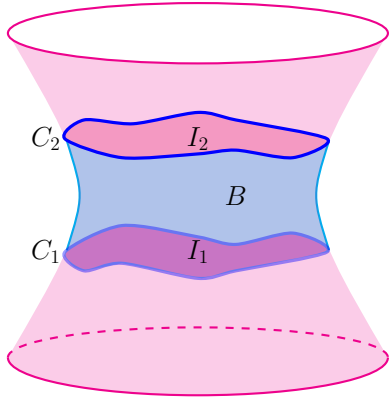
\begin{figure}[t]
 \centering
\begin{tikzpicture}[thick,scale=0.5, every node/.style={scale=0.5}]


\draw[cyan,thick] (6.45,1.45)..controls(6,0)..(6.45,-1.45)node{};
\draw[cyan,thick] (-.45,1.45)..controls(0,0)..(-.45,-1.45)node{};

\path[fill=magenta,opacity=0.2] (-2,4.3)..controls(.5,0)..(-2,-4.3)
 arc (180:360:5cm and 1cm)
 (8,-4.3)..controls(5.5,0)..(8,4.3)  arc (0:-180:5cm and 0.8cm)
 ;
\path[fill=cyan,opacity=0.3] (-.45,1.5)..controls(0,0)..(-0.52,-1.6)..controls(-.3,-1.9).. (0,-2)--(1,-1.8)--(3.1,-2.2)-- (4,-2)--(6.45,-1.5)..controls(6,0)..(6.45,1.5)--(-.45,1.5)
 ;

\draw[draw=magenta] (3,4.3) ellipse (5cm and 0.8cm);

\draw[magenta] (-2,-4.3) arc (180:360:5cm and 1cm)node{};
\draw[magenta,dashed] (-2,-4.3) arc (180:0:5cm and 0.8cm)node{};



\filldraw [fill=magenta,draw=blue,very thick,opacity=0.4]plot [smooth cycle] coordinates{(-0.5,-1.5) (0,-2)(1,-1.8)(3,-2.2) (4,-2)(6.4,-1.5)(5.5,-1)(4,-1.3) (3,-1.1)(1,-.8)};


\filldraw [fill=magenta!50!white,draw=blue,very thick,opacity=1]plot [smooth cycle] coordinates{(-0.5,1.5) (0,2)(1.3,1.9)(3,2.2) (4,2)(6.4,1.5)(5.5,1)(4,1.2) (3,1.1)(1,1)};
\node at (3,1.5) {\LARGE$I_2$};
\node at (3,-1.5) {\LARGE$I_1$};
\node at (-1,1.5) {\LARGE$C_2$};
\node at (-1,-1.5) {\LARGE$C_1$};
\node at (4,0) {\LARGE$B$};
\end{tikzpicture}
    \caption{$I_1, I_2$  depict constant de Sitter time ($\tau=const$) and blue-shaded region $B$ shows constant $\rho$ slice of the Mink$_d$ between the two constant $\tau$ regions. $C_1,C_2$ are boundaries of these constant time slices; these are two $S^{d-2}$ spheres corresponding to codimension 2 $\rho,\tau$-constant  surfaces in the Mink$_d$ region.}
    \label{boundary-in-dS-slicing}
\end{figure}

We show the large $\rho$ asymptotics of the Minkowski spacetime denoted as `the $dS_{d-1}$ boundary' by $B$ and the codimension 1 $\tau$-constant surfaces by $I$. Note that $B$ is a global $dS$ space. The boundaries of $dS_{d-1}$, indicated by $\partial B$ are then the intersection of $B$ with $I$-hypersurfaces. These are the $S^{d-2}$ surfaces at large $\rho$ and constant $\tau$ which are denoted as $C$ and depicted in Fig. \ref{boundary-in-dS-slicing}. 

\paragraph{Gauge fields and gauge parameters as differential forms.}

The $(p+1)$-form gauge fields of the $(d,p)$-theory will be denoted by $\bcA$ and the corresponding $(p+2)$-form field strength by $\bcF=\extd\bcA$:
\begin{align}
    \bcA&=\frac{1}{(p+1)!}\A_{\nu_0\cdots \nu_p}\extd x^{\nu_0}\wedge\cdots\extd x^{\nu_p}\,,\\
    \bcF&=\frac{1}{(p+2)!}\F_{\nu_0\cdots \nu_{p+1}}\extd x^{\nu_0}\wedge\cdots\extd x^{\nu_{p+1}}\,,
\end{align}
with\footnote{The anti-symmetric bracket is defined such that for a differential form $\omega$,
\eqs{
\omega_{[\mu_1\cdots\mu_n]}=\omega_{\mu_1\cdots\mu_n}\,,\qquad\;\;\; (\di\omega)_{\mu_1\cdots\mu_{n+1}}=(n+1)\partial_{[\mu_1}\omega_{\mu_2\cdots\mu_{n+1}]}\,.
}{}
}
\eq{\F_{\mu_0\cdots \mu_{p+1}}=(p+2)\partial_{[\mu_0}\A_{\mu_1\cdots\mu_{p+1}]}\,.}{}

\begin{table}[t]
\begin{center}

\begin{tabular}{ | m{2.0cm} || m{6.0cm}| m{5.7cm} | } 
\hline
\diagbox[width=6.0em]{\scriptsize{Space(time)}}{\scriptsize Forms}
 & \hspace{2cm}{Gauge fields} &  \hspace{1.4 cm}Gauge parameters\\ 
  \hline\hline
  \begin{center}
   Mink$_d$\end{center}
  & 
  \begin{center}$\bcA=\frac{1}{(p+1)!}\, \A_{\mu_0\cdots\mu_p}\,{\scriptstyle\di x^{\mu_0}\wedge\cdots\di x^{\mu_p}}\,$  \end{center}& \begin{center}
  $\boldsymbol{\Laa}=\frac{1}{p!}\,\Laa_{\mu_1\cdots\mu_p}\,{\scriptstyle\di x^{\mu_1}\wedge\cdots\di x^{\mu_p}}$ \end{center}\\ 
  \hline
  \begin{center}
 dS$_{d-1}$ \end{center}& 
\begin{center}
$\begin{array}{cl}
    \bA&= \frac{1}{(p+1)!}\,A_{a_0\cdots a_p}\,\scriptstyle\di x^{a_0}\wedge\cdots\di x^{a_p}\\
    \bA_\rho&= \frac{1}{p!} \,A_{\rho\, a_1\cdots a_p}\,{\scriptstyle \di x^{a_1}\wedge\cdots\di x^{a_p}}
\end{array}$\end{center}
& 
\begin{center}
$\begin{array}{cl}
    \boldsymbol{\laa}&=\frac{1}{p!}\,\laa_{a_1\cdots a_p}\,{\scriptstyle\di x^{a_1}\wedge\cdots\di x^{a_p}} \\
    \boldsymbol{\laa}_\rho&=\frac{1}{(p-1)!}\,\laa_{\rho\, a_2\cdots a_p}\,\scriptstyle\di x^{a_2}\wedge\cdots\di x^{a_{p}}
\end{array}$\end{center}
\\ 
  \hline\begin{center}
 S$^{d-2}$\end{center}& 
\begin{center}$\begin{array}{cl}
    \mathbf{\hat{A}}&=\frac{1}{(p+1)!}\,\hat{A}_{B_0\cdots B_p} \,\scriptstyle\di x^{B_0}\wedge\cdots\di x^{B_p}\\
    \mathbf{\hat{A}}_{\tau}&=\frac{1}{p!}\,\hat{A}_{\tau\, B_1\cdots B_p} \,\scriptstyle\di x^{B_1}\wedge\cdots\di x^{B_p}\\
    \mathbf{\hat{A}}_{\rho}&=\frac{1}{p!}\,\hat{A}_{\rho\, B_1\cdots B_p} \,\scriptstyle\di x^{B_1}\wedge\cdots\di x^{B_p}\\
    \mathbf{\hat{A}}_{\rho\tau}&=\frac{1}{(p-1)!}\,\hat{A}_{\rho\tau\,B_2\cdots B_p}\,\scriptstyle\di x^{B_2}\wedge\cdots\di x^{B_p}
\end{array}$\end{center}
& 
\begin{center}$\begin{array}{cl}
    \boldsymbol{\hat{\laa}}&=\frac{1}{p!}\,\hat{\laa}_{B_1\cdots B_p} \,\scriptstyle\di x^{B_1}\wedge\cdots\di x^{B_p} \\
    \boldsymbol{\hat \laa}_\tau&=\frac{1}{(p-1)!}\,{\hat \laa}_{\tau B_2\cdots B_p} \,\scriptstyle\di x^{B_2}\wedge\cdots\di x^{B_p} \\
    \boldsymbol{\hat \laa}_\rho&=\frac{1}{(p-1)!}\,{\hat \laa}_{\rho\,B_2\cdots B_p} \,\scriptstyle\di x^{B_2}\wedge\cdots\di x^{B_p}\\
    \boldsymbol{\hat \laa}_{\rho\tau}&=\frac{1}{(p-2)!}\,{\hat \laa}_{\rho \tau B_3\cdots B_p} \,\scriptstyle\di x^{B_3}\wedge\cdots\di x^{B_p}\\
\end{array}$\end{center}
\\ 
  \hline
\end{tabular}
\end{center}
    \caption{Summary of notations for forms. Bold letters $\bcA$($\boldsymbol{\Laa}$), $\mathbf{A}$($\boldsymbol{\laa}$) and $\hat\bA$($\boldsymbol{\hat \laa}$) stand for abstract gauge fields (parameters) as differential forms 
    on Mink$_d$, $dS_{d-1}$ and $S^{d-2}$ respectively.}
    \label{tab:notation}
\end{table}

{Throughout this paper, we will be dealing with forms on the Minkowski space $\mathbb{R}^{d-1,1}$, the de Sitter $dS_{d-1}$ and the sphere $S^{d-2}$. The components of forms in any case will be indicated by an assigned font as shown in Table \ref{tab:notation} and the associated forms with the same font but boldfaced.} 
We select an asymptotic $\rho$-expansion for the gauge fields and discriminate the leading components by uppercase Latin font,
 \eq{
 \A_{\mu_0\cdots\mu_p}(\rho,x^a)=A_{\mu_0\cdots\mu_p}(x^a)\rho^n+\text{subleading}\,,
 }{asympA}
 for an appropriately chosen fall-off power $n$. Note that $n$ may be different for different components. For our case, if the fall-off for components involving the $\rho$ direction is $n$, for the other components along $dS$ (not involving $\rho$) it is $n+1$; see section \ref{se:boundary conditions}. 
The $A_{\mu_0\cdots\mu_p}(x^a)$ fields, which have dependence only on $x^a$ , may be viewed as form fields on the de Sitter of unit radius. 
That is, one may decompose the $\mu_i$ indices into $\rho, x^a$
and hence $A_{\mu_0\cdots\mu_p}(x^a)$ yields  a $(p+1)$-form $\bA$ and a $p$-form $\bA_{\rho}$ on $dS_{d-1}$;

\begin{align}\label{Arho-form}
    \bA_{\rho}&\equiv\rho^{-n}\,\mathbf{n}\cdot \bcA\Big|_{\rho\to\infty}=\frac{1}{p!} A_{\rho\, a_1\cdots a_p}dx^{a_1}\wedge\cdots dx^{a_p}\,,\\
\bA&\equiv \rho^{-(n+1)}\Big( \bcA-\di\rho\wedge\mathbf{n}\cdot\bcA\Big)\Big|_{\rho\to\infty}=\frac{1}{(p+1)!} A_{a_0\cdots a_p}dx^{a_0}\wedge\cdots dx^{a_p}\,,
\end{align}
where $\mathbf{n}=\partial_\rho$ is the unit normal vector to constant $\rho$ surfaces, i.e. the de Sitter slices.
The above should of course be computed at $\rho\to\infty$ where the `boundary' $B$ is defined (\emph{cf.} Fig. \ref{boundary-in-dS-slicing}).  We will raise the indices on $\mathbf{A}$ and $\mathbf{A}_\rho$ by $h_{ab}$ the unit de Sitter metric.
 In our analysis of the charges we will work with forms on the ($t,r$)-constant/($\rho,\tau$)-constant  sections of the spacetime that is $S^{d-2}$. 

To distinguish forms on the sphere $S^{d-2}$ from those on the de Sitter $dS_{d-1}$ we will denote the former by hatted uppercase Latin letters, as shown in the Table \ref{tab:notation}. In particular, the $(p+1)$-form gauge field of the bulk is decomposed into one $(p+1)$-form, two $p$-forms and one $(p-1)$-form on $S^{d-2}$; 
\begin{align}
{\hat\bA}&\equiv\bA-\di\tau\wedge{\hat \bA}_{\tau}\,,\;\;\qquad  \;\; {\hat \bA}_{\tau}\equiv\boldsymbol{\tau}\cdot \bA\,,\\
{\hat \bA}_{\rho}&\equiv\bA_\rho-\di\tau\wedge{\hat \bA}_{\rho\tau}\,,\;\qquad {\hat \bA}_{\rho\tau}\equiv\boldsymbol{\tau}\cdot \bA_\rho\,,\label{Atau-form}
\end{align}

where $\boldsymbol{\tau}=\partial_\tau$ is the normal vector to constant $\tau$ surfaces of the de Sitter slices, normalized with respect to $h_{ab}$. Similarly the $p$-from gauge parameter $\boldsymbol{\Lambda}$ of the bulk is reduced to one $p$-form, two $(p-1)$-forms and one $(p-2)$-form on the $S^{d-2}$, see Table \ref{tab:notation}.

\paragraph{Hodge decomposition of forms on sphere.}
Our expressions for the charges are given by codimension 2 surface integrals of a $(2p+2)$-from on the $S^{2p+2}$. The integrands are composition of two $(p+1)$-forms or a $p$ and a $(p+2)$-form on the sphere. The sphere is a compact Riemannian manifold, and hence one can use the Hodge theorem to decompose these forms on it (see appendix \ref{Ap-diff} for more discussions). In particular, for a $p$-form gauge parameter we have,
\eq{\hat{\boldsymbol{\laa}}=\hat{\boldsymbol{\laa}}^{\textrm{\tiny exact}}+\hat{\boldsymbol{\laa}}^{\textrm{\tiny coexact}}
+\hat{\boldsymbol{\laa}}^{\textrm{\tiny harmonic}}\,,
}{exact-coexact-decom}
where
i) $\hat{\boldsymbol{\laa}}^{\textrm{\tiny harmonic}}$ is annihilated by the Laplace-Beltrami operator $\Delta=\di\di^\dagger+\di^\dagger \di$,
ii)
$\hat{\boldsymbol{\laa}}^{\textrm{\tiny exact}}=\di{\hat{\boldsymbol\phi}}$ is the curl of a lower rank form, and
iii) $\hat{\boldsymbol{\laa}}^{\textrm{\tiny coexact}}=\di^\dagger{\hat{\boldsymbol\psi}}$ is the divergence of a higher rank form. 

We emphasize that in this work, only forms on $S^{d-2}$ are subject to this decomposition. As a side remark, {harmonic forms on sphere exist only for functions and top forms, that is, if
 $p= 0$ or $d-2$. As a result, except for the $(4,0)$-form theory (i.e. the $4d$ Maxwell theory), we only deal with exact and coexact gauge parameter forms.
 
Finally, we point out that we denote the Hodge dual of a form $\mathbf{X}$ by $\star\mathbf{X}$, where the star operation  is understood to be on Minkowski, de Sitter or the sphere depending on the type of the font used for the form as defined on these space(-times). {Similarly, the exterior calculus on all three spaces is carried out by the same symbols $\di,\,\di^\dagger,$ and $\Delta$.}


\section{Basic setup }\label{se:action}

In this section we introduce our theory through the action principle and by specifying 
the boundary conditions.
\subsection{Action for \emph{(d,p)}-form gauge field theory}
We start with the action of the $p$-form gauge theory in the $d$-dimensional Minkowski spacetime $\M_d$, denoted as ($d,p$)-theory,\footnote{
One may of course consider other gauge invariant actions e.g. $p$-form  Chern-Simons theory \cite{Edelstein:2008ry,Bekaert:2002eq,Bunster:2011qp} or Born-Infeld theory \cite{Kalb:1974yc,Chruscinski:2000zm}.}
\eq{S=S_0+S_b=-\int_{{\mathcal M}_d} \Big(\tfrac12\extd {\bcA}\wedge \star\extd {\bcA} -(-1)^p {\bcA} \wedge \star{\bcJ}\Big)+\int_{{\partial\mathcal M}_d} {\cal L}_b\,,}{pformaction}
 which is the generalization of the Maxwell theory for $(p+1)$-form gauge fields $\bcA$ and the $(p+1)$-form currents $\bcJ$. The last term is a potential boundary term defined by an integral over the boundary of $\M_d$,  $\partial\M_d$. In the de Sitter slicing $\partial{\cal M}_d$ is the union of regions $I_1,I_2, B$ depicted in Fig. \ref{boundary-in-dS-slicing}. 

Variation of the action \eqref{pformaction} leads to,
\eq{\delta S=(-1)^{p+1}\int_{\mathcal M} \delta \bcA\wedge\Big( \extd\star\extd \bcA - \star \bcJ\Big)+\int_{{\partial\mathcal M}_d} (\delta {\cal L}_b-\delta \bcA\wedge\star\extd \bcA)\,.}{actionprinc}
Action principle yields equations of motion
\eq{ \extd\star\extd \bcA =\extd \star\bcF=\star \bcJ, \qquad\text{or}\qquad \nabla_{\alpha}\F^{\mu_0\cdots\mu_p\alpha}=\mathcal{J}^{\mu_0\cdots\mu_p},}{eom}
provided that our boundary conditions ensure
\eq{(\delta \bcA\wedge\star\extd \bcA-\delta {\cal L}_b)|_{\partial{\mathcal M}_d}=0.}{bdry-term=0}
We will discuss this latter condition in more detail in  section \ref{variational-section}.

\subsection{Gauge symmetry}

Integrability of \eqref{eom} implies $\extd \star\bcJ=0$, which in turn yields gauge invariance of the action \eqref{pformaction}, if ${\cal L}_b$ is also invariant. Explicitly, the action \eqref{pformaction} is invariant under
\eq{\bcA\to \bcA+\extd\boldsymbol{\Lambda},}{gauge-transf}
where $\boldsymbol{\Lambda}$ is a $p$-form. 
To analyze the theory and the associated conserved charges we need to fix the gauge freedom \eqref{gauge-transf}. The gauge fixing condition for a $(d,p)$-form theory is generically to set a $p$-form combination of  form fields or their first derivatives equal to zero. A convenient choice which is also \emph{Lorentz covariant} is the Lorenz gauge
\eq{\di^\dagger\bcA=0,\qquad\text{or}\qquad-\nabla_\alpha\A^{\alpha\mu_1\cdots\mu_p}=0}{Lorenz-gauge}
where $\di^\dagger=(-1)^{d(p+1)}\star\di\star$ is the co-differential operator on Minkowski space (we use the same notation for de Sitter and sphere discussed in the appendix \ref{Ap-diff}). In this gauge the equations of motion \eqref{eom} are

\eq{\Delta \bcA=(-1)^p\boldsymbol{\mathcal{J}}\,,\qquad \Delta=\extd\,\extd^\dagger+\extd^\dagger\extd \,.}{eomLG}
The gauge condition \eqref{Lorenz-gauge}  separates into two sets of conditions in the de Sitter slicing
\eq{
D_{a_0}\A^{a_0a_1\cdots a_p}+\left(\frac{d-1}{\rho}+\partial_\rho\right)\A^{\rho\,a_1\cdots a_p}=0\,,
 \qquad D^{a_1}\A_{\rho a_1\cdots a_p}=0\,.}{LG}

Having fixed the Lorenz gauge, we remain with $p$-form \emph{residual gauge symmetries}: The gauge fixing condition \eqref{Lorenz-gauge} still allows gauge transformations of the form \eqref{gauge-transf} where the $p$-form $\boldsymbol{\Lambda}$  satisfies
\eq{
\di^\dagger\di\boldsymbol{\Laa}=0\,.}{residual0}
Nevertheless, \eqref{residual0} indicates that we can still define $\boldsymbol{\Lambda}$ up to an exact form. This freedom may be fixed by setting an extra condition on $\boldsymbol{\Lambda}$ analogous to the Lorenz gauge \eqref{Lorenz-gauge} on the gauge field $\bcA$. We will return to this point at section \ref{Charge-classification-sec}.
\paragraph{Propagating degrees of freedom.}
The $(p+1)$-form gauge field in $d$ spacetime dimensions has $\binom{d}{p+1}$ independent components. However, gauge symmetry implies that only transverse modes of the form field are propagating. The $p$-form gauge parameter enables us to remove $\binom{d-1}{p}$ components. This could be done e.g.  by imposing the covariant gauge condition \eqref{Lorenz-gauge}. The residual $p$-form gauge parameters satisfying \eqref{residual0} gauge away another $\binom{d-2}{p}$ components. The total number of degrees of freedom turns out to be,
\eq{\binom{d}{p+1}-\binom{d-1}{p}-\binom{d-2}{p}=\binom{d-2}{p+1},}{number-of-dof}
where the last term on the LHS in \eqref{number-of-dof} is the contribution of the residual degrees of freedom. In the appendix \ref{Hamiltonian} we have presented a more precise counting of propagating degrees of freedom in Hamiltonian formalism by counting first-class constraints and reducibility identities. 

The number of degrees of freedom \eqref{number-of-dof} can be intuitively understood as the number of independent components of the $(p+1)$-form in the transverse $d-2$ dimensional space. That is, the propagating modes can be explicitly described through solution to equations of motion for a radiation in the transverse gauge:
\begin{equation}
\A_{\mu_0\cdots\mu_p}= \epsilon_{\mu_0\cdots\mu_p} e^{ik_\mu x^\mu},\qquad k^{\mu}k_{\mu}=0,\ k^{\mu_i} \epsilon_{\mu_0\cdots\mu_i\cdots\mu_p}=0,\ i=0,\cdots,p.    
\end{equation}

\subsection{Zero mode charges of \emph{(d,p)}-form theory}\label{zero-mode}

Consistency of field equations \eqref{eom} requires the current $\bcJ$ be coclosed $\extd \star\bcJ=0$, giving rise to a conserved quantity \cite{Henneaux:1986ht},
\eq{Q=\int_{{\mathcal M}_{d-p-1}}\star\bcJ\,,}{}
where $\M_{d-p-1}$ is the spacelike surface orthogonal to the worldvolume of $p$-brane source. If there are several parallel $p$-branes in space, $\M_{d-p-1}$ intersects each one at a point, and the charge will be the total number of branes with sign $+$ or $-$ for each brane according to its orientation. For example, a couple of parallel branes with opposite orientations have zero total charge. Using the Stokes' theorem, the charge \emph{density} can be expressed as a Gauss's law in the orthogonal space $\M_{d-p-1}$ whose boundary is an $S^{d-p-2}$
\eq{
Q=\int _{S^{d-p-2}}\star \bcF\,.
}{gauss p-form}
The quantity \eqref{gauss p-form} counts the net electric charge of the parallel branes by integration on orthogonal space\footnote{\eqref{gauss p-form} differs from that in \cite{Henneaux:1986ht} duo to different  sign conventions at the action level.
}. The conserved charge \eqref{gauss p-form} may be directly related to the global part of gauge transformations \eqref{gauge-transf} i.e. gauge transformations with  $\extd\boldsymbol{\Lambda}=0$, via the standard Noether's theorem, 
\eq{Q_{\boldsymbol{\Lambda}}[\bcF]=\int_{S^{d-2}}\boldsymbol{\Lambda}\wedge{}\star \bcF\,.}{Noether-charge}
Taking the gauge parameter to be proportional to the volume-form of the $p$-brane, namely, $\boldsymbol{\Lambda}\propto dx^1\wedge\cdots dx^p$, then \eqref{Noether-charge} reduces to \eqref{gauss p-form}.\footnote{To do the calculation for this case, one could work in cylindrical coordinates aligned with the brane. The integration on $p$-brane directions is then trivial and gives an infinite multiplicative factor which we drop from charge value.}

One can make the above discussion more general and systematic, allowing arbitrary relative alignments for the branes.
 There is a set of gauge transformations, the exact symmetries in the language of \cite{Hajian:2015xlp}, that keeps any gauge field  $\bcA$ intact; 
 satisfying $\extd\boldsymbol{\Lambda}=0$. 
In the $p=0$ case of Maxwell theory, where $\boldsymbol{\Lambda}$ is a 0-form, the only solution to $\extd\boldsymbol{\Lambda}=0$ is constant $\boldsymbol{\Lambda}$ which produces the electric charge. Also the case of $p=d-2$ has been studied in \cite{Chernyavsky:2017xwm}. For the $0<p<d-2$ case,
 we define the \emph{zero-mode charges} of $(d,p)$-form theory to be generated  by closed $p$-form gauge parameters $\boldsymbol{\Lambda}_{B_1 \cdots B_p} (\hat{x})$,  defined on the asymptotic $(d-2)$ sphere with no $r$- and $t$- dependence. Such forms are exact $\boldsymbol{\Lambda}=\di\boldsymbol{\epsilon}$ and \eqref{Noether-charge} becomes
\eqs{
Q[\boldsymbol{\epsilon}]=(-1)^p
\int_{S^{d-2}}\boldsymbol{\epsilon}\wedge\di{}\star \bcF= (-1)^p\int_{S^{d-2}}\boldsymbol{\epsilon}\wedge\star \bcJ\,,
}
 where field equations have been used in the last equality.\footnote{Note that field equations involve exterior derivative $\di$ on Minkowski space, not  on the sphere. Nonetheless, when the equation is pulled back to the sphere at $r$-, $t$- constant, the exterior derivative reduces to that of the sphere.
 } In terms of components, 
 \eqs{
 Q[\epsilon]=\frac{(-1)^p}{(p-1)!}\int_{S^{d-2}}d\Omega_{d-2}\,\epsilon_{B_2\cdots B_{p}}{\mathcal{J}}^{trB_2\cdots B_{p}}.\label{zero charge general p}
 }
For \eqref{zero charge general p} to be non-vanishing, the source ${\mathcal{J}}$ must  extend to infinity, like an infinite string or a planar brane. For example, an infinite string induces a couple of points on the celestial sphere with opposite signs according to its orientation. As a result, objects like closed loops (which do not extend to infinity) have  zero charge. Increasing the form rank by one, a planar source induces a great circle on the celestial sphere with definite orientation for its tangent vector.

The zero-mode charge is only sensitive to the \emph{asymptotic alignment} of the brane and not to its shape or velocity inside the bulk. If the asymptotic alignment of the source is time-independent (which is physically reasonable), the zero-mode charge \eqref{zero charge general p} is clearly conserved.
To compute the zero-mode charge explicitly, the simplest yet non-trivial example is $(6,1)$-form theory, where the sources are strings and $\epsilon$ is a 0-form (a function) on 4-sphere. The charge can be written in terms of the sources
\eqs{
Q[\epsilon]=-\int_{S^4}d\Omega_{d-2}\,\epsilon\, \mathcal{J}^{tr}(\hat{x}).\label{zero-mode charge}
}
To be explicit, let's consider a number of $n$ straight strings aligned at $\hat{s}_n$ directions. The source will be
\eqs{
\mathcal{J}^{tr}=\sum_nq_n\hat{x}\cdot \hat{s}_n
[\de^4(\hat{s}_n)+\de^4(-\hat{s}_n)]/\sqrt{\text{det}\mathscr{G}_{AB}},
}
and the charge is
\eqs{
Q[\epsilon]=-\sum_nq_n\Big[\epsilon(\hat{s}_n)-\epsilon(-\hat{s}_n)\Big].
}
This is the intuitively expected result. For every function $\epsilon(x^A)$ on the 4-sphere there is one charge and the whole set has the information of the alignment of all strings and their charge $q_n$. For another approach to zero-mode $p$-form charges see \cite{Compere:2007vx}.

\subsection{Boundary conditions}\label{se:boundary conditions}

In order to fully introduce the theory we have to specify the boundary conditions on the dynamical fields. This together with the equations of motion determines the space of field configurations that defines the theory. In particular one imposes a set of boundary conditions that determine how the fields decline  at infinity. Sometimes, specific \emph{gauge} conditions are also set to narrow the space of functions under consideration  further. 

\paragraph{Radiation and Coulomb fall-off behaviors.} There are usually two physically relevant fall-off behaviors, the Coulomb and radiation fall-offs.
For the Coulomb fall-off behavior, let us consider field strengths $\bcF$ that represent the `electric charges' of the theory. In $p=0$ case, they are electric monopoles moving freely in space, while in generic $(d,p)$-form theories, the sources are extended $p$-branes. A static $p$-brane in Minkowski spacetime, extended in $x_1$ to $x_p$ directions is described by the source
\eq{\bcJ\propto\di x^0\wedge\cdots\di x^p.}{}
 Solving field equations \eqref{eom} gives
\begin{equation}
{\cal A}_{01\,\cdots \,p}\propto\frac{1}{\ell^{d-p-3},}\label{cartesian b.c}
\end{equation}
where $\ell^2=\sum_{k=p+1}^{d-1}(x^k)^2$ is the orthogonal distance to the brane. For $p=0$, this is the familiar $1/r^{d-3}$ behavior of the electric field and is hence called Coulomb fall-off behavior. Boosting the brane gives rise to purely spatial magnetic components of the field strength with the same fall-off. 

The radiation fall-off behavior corresponds to intensity of (black-body) radiation ${\cal E}$ of a localized gas of `$(p+1)$-form photons' in $d$ dimensions. This is given by ${\cal E}\propto 1/r^{d-2}$. This energy is carried by the radiation $(p+1)$-form field with temporal component ${\cal A}_u$ (in the standard Bondi frame at null infinity) such that ${\cal E}\propto (\partial {\cal A}_u)^2$, yielding the radiation fall-off behavior ${\cal A}_u\sim 1/r^{\frac{d-2}{2}}$ \cite{Ortaggio:2014ipa, Campoleoni:2017qot}.

For $d>2p+4$, Coulomb field falls off faster than radiation and the converse is true for $d<2p+4$. In $d=2p+4$, which we will be interested in, both radiation and Coulomb fields fall off in the same rate and hence the traces of passage of $(p+1)$-form radiation can be recorded in the associated $(p+1)$-form charges\footnote{\label{scaling footnote}$d=2p+4$ is also the dimension in which our $(d,p)$-form theory exhibits conformal symmetry \cite{Deser:1994ca,Raj:2016zjp}. The scaling part of this symmetry may be readily seen: let us start with the geometric object $\bA$ $(p+1)$-form and require that it is invariant under scaling. This means the form field components ${\cal A}_{\mu_0\cdots\mu_p}$ should scale as $\lambda^{p+1}$ if we scale $x^\mu\to \lambda^{-1} x^\mu$. With this scaling the Lagrangian density $\sqrt{g}|d{\cal A}|^2$ scales by $\lambda^{-d}\lambda^{2(p+2)}$. Scale invariance of the action hence yields $d=2p+4$. This scale invariance for our theory is enhanced to the $2p+4$ dimensional conformal symmetry \cite{Deser:1994ca}.}, leading to $p$-form memory effect.

\paragraph{Fall-off behavior in de Sitter slicing.} According to the previous discussion, one can verify that in the de Sitter slicing introduced in section \ref{sse:notation} 
the fall-off behavior of field strength associated with a $p$-brane source for any $p$ and $d$  is\footnote{Note that $F$ with lower indices has $\rho^{-(d-p-1)}$ fall-off. However, for convenience and later use in \eqref{bnyc1} we have presented large $\rho$ behavior of $F$ with upper indices.}
\begin{subequations}
\begin{align}\label{bnyc1}
 \F^{\rho a_0\cdots a_p}=&F^{\rho a_0\cdots a_p}(x^b)\,\rho^{-d+1}+\ordr{-d},\\
\F^{a_0\cdots a_{p+1}}=&F^{a_0\cdots a_{p+1}}(x^b)\,\rho^{-d}+\ordr{-d-1}\,.  \label{bnyc0}
\end{align}\label{bnyc10}
\end{subequations}
The boundary conditions on gauge fields compatible with the above are
\begin{subequations}
\begin{align}
\A^{\rho a_1\cdots a_p}&=A^{\rho a_1\cdots a_p}(x^b)\,\rho^{-d+3}+\ordr{-d+2},\label{bnyc2}\\
\A^{a_0\cdots a_{p}}&=A^{a_0\cdots a_{p}}(x^b)\,\rho^{-d+2}+\ordr{-d+1}\,.\label{bnyc3}
\end{align}\label{bnyc23}
\end{subequations}
These boundary conditions are preserved by the following (residual) gauge transformations;
\begin{subequations}
\begin{align}
\Lambda^{\rho a_2\cdots a_{p}}&=\lambda^{\rho a_2\cdots a_{p}}(x^b)\,\rho^{-d+5}+\ordr{-d+4},\label{bpgt2}\\
\Lambda^{a_1\cdots a_{p}}&=\lambda^{a_1\cdots a_{p}}(x^b)\,\rho^{-d+4}+\ordr{-d+3}\,.\label{bpgt3}
\end{align}
\end{subequations}

We note that \eqref{bnyc2} and \eqref{bnyc3} are not the only possibilities which follow from \eqref{bnyc1}. In principle we could have chosen a weaker fall-off behavior for allowed gauge transformations than the background gauge fields $\A_{\mu_0\cdots\mu_p}$. Such `leading' gauge transformations do not yield a finite charge (as we compute in section \ref{section-3}) and hence we will not study them in this work.  

\subsection{Action principle and the boundary term}\label{variational-section}

On a $d$-dimensional Lorentzian globally hyperbolic manifold $\M$, let $S[\Phi(t,\x)]$ be a functional of the set of generic fields $\Phi(x^\mu)$ on $\M$ with at most two time derivatives. For fixed initial and final data;
\begin{equation}
\Phi(t,\x)\big|_{t=i}=\Phi_i(\x)\, ,\qquad
\Phi(t,\x)\big|_{t=f}=\Phi_f(\x)\,,\label{initial}
\end{equation}
the classical trajectory $\Phi_{cl}(t,\x)$ is defined as the solution of
\begin{equation}
   E(\Phi)\equiv \frac{\delta S}{\delta \Phi(t,\x)}=0\,.
\end{equation}
The action $S$ is said to have a well-defined action principle if
\be\label{variation-principle-def}
\hspace*{-5mm}\delta S=\int_\M E(\Phi)\delta \Phi+\int_{I_2}\I_0(\delta \Phi, \Phi)- \int_{I_1}\I_0(\delta \Phi, \Phi)+\int_{\partial I_2}\I_b(\delta \Phi, \Phi)-\int_{\partial I_1}\I_b(\delta \Phi, \Phi),
\ee
for generic field variations $\delta \Phi=\delta \Phi(t,\x)$. Here $I_1,I_2$ denote constant time slices of $\M$ at $t_i,\,t_f$ respectively, and ${\partial I_1}, {\partial I_2}$ are the (spacelike) boundaries of the constant time slices.
That is,  variation of an  action with well-defined action principle is vanishing on-shell upon a suitable fixation of initial and final conditions under generic field variations.

Given the definition above, a generic action $S_0$ with a prescribed boundary condition on fields, may fail to obey a well-defined action principle due to appearance of a boundary term $\B$ on the time-like boundary $B$:
\be\label{gen variation}
\delta S_0=\int_\M E(\Phi)\delta \Phi+\int_{I_2}\I_0(\delta \Phi, \Phi)- \int_{I_1}\I_0(\delta \Phi, \Phi)+\int_{B}\B(\delta \Phi, \Phi),
\ee
where $B$ is the timelike boundary  ($\rho=\rho_0=$constant in de Sitter slicing), and $\I_0$ is the initial/final term integrated on the initial spacelike constant time (constant $\tau$ in de Sitter slicing) surfaces  $I_1$ or the final one $I_2$ such that, $\partial \mathcal{M}=B\cup I_1\cup I_2$; \emph{cf.} Fig. \ref{boundary-in-dS-slicing}. In these cases we supplement $S_0$ with a suitable boundary term $S_b=\int_B {\cal L}_b$ such that, off-shell,
\be
\delta S_b=-\int_{B}\B(\delta \Phi, \Phi)+ \int_{\partial I_2}\I_b(\delta \Phi, \Phi)-\int_{\partial I_1}\I_b(\delta \Phi, \Phi).
\ee
where $\B$ is the boundary term of $S_0$ in \eqref{gen variation}, integrated on the timelike boundary.  In consequence, $S_0+S_b$ defines a  well-defined boundary value problem in the sense that the first variation of the action  under all field variations that preserve our boundary conditions on $B$ is of the form \eqref{variation-principle-def}. 

To see how this works in practice and how one fixes  $S_b$, consider a generic variation of $S_0$ in the $(d,p)$-action \eqref{pformaction} and plug the boundary conditions \eqref{bnyc10}-\eqref{bnyc23} in \eqref{actionprinc}, we find,
\begin{align}\label{boundary remain}
   \B\text{-term}&=-\frac{\rho_0^{d-1}}{(p+1)!}\int_{B}\sqrt{-h}\,n_{\alpha}\, \delta \A_{\mu_0\cdots \mu_p} \F^{\alpha\, \mu_0\cdots \mu_p}\\
            &=-\frac{\rho_0^{2p+4-d}}{(p+1)!}\int_{B}\sqrt{-h}\, \delta A^{a_0\cdots a_p} \left[(2p+4-d)A_{a_0\cdots a_p}-(p+1)\partial_{a_0} A_{\rho\,a_1\cdots a_p}\right]\nn.
\end{align}
where $\mathbf{n}=\partial_\rho$ is the unit vector normal to the $\rho=\rho_0$  hypersurface.
As we see for large $\rho_0$, this term vanishes for $d>2p+4$ case (where the Coulomb fall-off is faster than that of radiation, \emph{cf.} our discussions in previous section), while for $d<2p+4$ case the boundary term blows up unless field variations identically vanish at the boundary. The case of $d=2p+4$, which as argued in previous sections is the case we are focusing on in this work, is the case where the boundary term remains finite and we need to add a non-trivial boundary term for any given variation, as we will discuss below.

For $d=2p+4$ case the first term in \eqref{boundary remain} drops and the remaining one is $\ordr{0}$. This boundary term is generically non-zero for our boundary conditions and spoils our boundary-value problem. However, we could get rid of this term by fixing a gauge condition and introducing the boundary term $S_b$. To this end we rewrite the Lorenz gauge fixing condition \eqref{LG} to the leading term in $\rho$
\begin{subequations}
\begin{align}
D^{a_0}A_{a_0a_1\cdots a_p}+2A_{\rho\,a_1\cdots a_p}&=0\,,\label{Lorentz gauge split}\\
D^{a_1}A_{\rho \,a_1\cdots a_p}&=0\,.\label{LG-2}
\end{align}
\end{subequations}
Using \eqref{Lorentz gauge split}   the $\B$-term takes the following form,
\begin{align}\label{dBterm}
        \B\text{-term} =-\frac{1}{p!}\int_{B}\sqrt{-h}\,\delta\left( A^{\rho}\cdot A_{\rho}\right)_p-
        \frac{1}{p!}\int_{\partial B}\sqrt{\mathscr{G}}\,\tau_b\left(\delta A^b\cdot A^{\rho}\right)_p\,,
        \end{align}
where $\boldsymbol{\tau}=\partial_\tau$ is the `outward-pointing' normal vector to the $\tau$-constant  hypersurfaces in  unit de Sitter normalized as $h^{ab}\tau_a\tau_b=-1$, and we have introduced the notation,
\eq{\left(A\cdot B\right)_p\equiv A_{a_1\cdots a_p} B^{ a_1\cdots a_p}\,.}{AdotB}
The second term in \eqref{dBterm} is an integration on  $\partial B$ which is the intersection of $B$ with $I_1$ and $I_2$ in \eqref{gen variation} (see Fig. \ref{boundary-in-dS-slicing}). 
In a well-defined initial value problem, one fixes the initial and final values of the variable, so this term vanishes. Since the $\B$-term is a total variation, it is sufficient to identify the boundary term $S_b$ in the original action \eqref{pformaction} as,
 \begin{equation}\label{Bterm0}
 S_b =-\int_{B}\, \bA_{\rho}\wedge \star \bA_{\rho}\,.
 \end{equation}
where $\bA_\rho=\frac{1}{p!}A_{\rho\,a_1\cdots a_p}dx^{a_1}\wedge\cdots dx^{a_p}$ is the boundary $p$-form while the wedge and the Hodge duality is defined on $B$  --- see section \ref{sse:notation}. 
This ensures to have a well-defined action principle for the $(2p+4, p)$-form theory with boundary conditions \eqref{bnyc10}-\eqref{bnyc23} in the Lorenz gauge \eqref{Lorenz-gauge}.\footnote{Such a boundary term appears also in the context of $AdS_2$ holography \cite{Castro:2008ms}.} 

\paragraph{Zero-modes charges and action principle.} In our action principle analysis above we required the variation of action to be zero on-shell for generic field variations $\delta \A$ with prescribed fall-off behavior \eqref{bnyc10}-\eqref{bnyc23}. One should, however, note that there could be other restrictions on the physically allowed/relevant variations. One particular case we discuss here concerns the case where the $p$-brane sources are on. One may hence consider field variations over the setup with a given fixed source content. In statistical mechanical systems this is analogous to studying systems in sectors with a given 
chemical potential. In the $(d,p)$-form theories, one may observe that in the presence of $p$-brane sources, $\bar{F}_{\tau \rho \,B_1\cdots B_p}$ (the field produced by static sources with specific zero-mode charges) is playing the role of a fixed chemical potential on $S^{d-2}$, and 
on-shell we generically have;
\eq{\delta S\sim\int_B \delta { A}_{\tau B_1\cdots B_p}\bar{F}^{\tau \rho\,B_1\cdots B_p}\,.}{varbackground} 
Since $\bar{F}_\rho$ is fixed, one may get rid of this term either by restricting the boundary conditions on $\delta { A}_\tau$ such that the contribution above is zero or we may add the suitable boundary term which is minus the field-space integration of the above contribution. 
\paragraph{Invariance  under residual gauge transformation.} Boundary term \eqref{Bterm0} is obtained by Lorenz gauge fixing which leaves us with the residual gauge transformations, generated by ${\boldsymbol{\Lambda}}$ satisfying \eqref{residual0}. For consistency of our analysis, especially the computation of charges discussed in the next section,  one should hence make sure that this boundary term respects this residual gauge symmetry. Gauge transformations \eqref{gauge-transf} at leading order are
\begin{subequations}
\begin{align}
    \delta_\laa A_{\rho\,a_1\cdots a_p}&=-p\,\partial_{[a_1}\laa_{|\rho|\,a_2\cdots a_p]}\qquad\text{or}\qquad \delta_\laa\mathbf{A}_\rho=-\di\boldsymbol{\laa}_\rho\,,\label{gauge-transf-leading1}\\
\delta_\laa A_{b\,a_1\cdots a_p}&=(p+1)\,\partial_{[b}\laa_{a_1\cdots a_p]}\qquad\text{or}\qquad \delta_\laa\mathbf{A}_b=(\di\boldsymbol{\laa})_b\,.\label{gauge-transf-leading2}
\end{align}\label{residualdS}
\end{subequations}
Obviously, if $p\geq 1$, the boundary term \eqref{Bterm0} is not invariant under \eqref{gauge-transf-leading1}. However, as we argue here, the extra terms can be cast into integrals on the future and past codimension 2 boundaries of the de Sitter space using Lorenz condition \eqref{LG-2}. 
Gauge transformation of the boundary term is explicitly
  \begin{equation}\label{Bterm1}
  \delta_\lambda S_b =2(-1)^{p}\int_{ B} \boldsymbol{\laa}_\rho\wedge \extd\star\bA_\rho +2\int_{{\partial B}} \boldsymbol{\laa}_\rho\wedge\star\bA_\rho\,.
  \end{equation}
 The first integral is zero as a consequence of the Lorenz gauge condition \eqref{LG-2} and the second integral is at boundaries of the de Sitter space where all large gauge transformations also act. Although the boundary term  \eqref{Bterm0} is only defined on the timelike segment $B$ of the boundary $\partial \mathcal{M}$, we can simply extend it to the whole boundary by fixing initial and final data on $I_{1,2}$. 
The improved action of our $(2p+4,p)$-form theory is thus,
\begin{equation}\label{improvedaction}
    S=-\frac{1}{2(p+2)!}\int_{\mathcal M}\sqrt{-g}\,\F_{\mu_0\cdots\mu_{p+1}}^2-\frac{1}{p!}\int_{\partial {\mathcal M}}\sqrt{-h}\,\left(A^{\rho }\cdot A_{\rho }\right)_p\,.
\end{equation}

Although we have fixed the boundary fall-off behavior of our fields, we may need  further restrictions on the fields to make sure that the on-shell action (i.e. the boundary term) is a finite quantity for given consistent initial and final conditions. Since field equations for $\bA_{\rho}$ are of second degree in de Sitter time $\tau$, for each angular mode there are two functions of $\tau$.

\section{Conserved charges of residual gauge symmetries}\label{section-3}

Setting the stage in the previous section, we now turn to computing conserved asymptotic surface charges associated with the boundary condition preserving residual gauge symmetries of the $(2p+4,p)$-form theory.  Here we use an extension of the Noether theorem and our analysis is based on a simple and direct manipulation of the action.  Let us review how symmetries of generic {gauge} theories give rise to conserved charges {in flat spacetime}. Consider a generic variation of an action $S$, already including the boundary terms required for having a well-defined action principle, e.g. like the one in 
\eqref{improvedaction}. The variation of this action for general field variations has the form \eqref{variation-principle-def}.
A variation $\delta_\epsilon\Phi$ of the fields, generated by a continuous parameter $\epsilon$ is a symmetry, if it leaves the action invariant off-shell, up to possible terms over constant initial and final time slices \cite{Henneaux:1992ig}, i.e. 
\be\label{symmetry-def}
\begin{split}
\delta_\epsilon S_0&=\int_{I_2}{\cal K}_0 (\delta_\epsilon\Phi,\Phi)-\int_{I_1}{\cal K}_0 (\delta_\epsilon\Phi,\Phi),\\
\delta_\epsilon S_b&=\int_{\partial I_2}{\cal K}_b (\delta_\epsilon\Phi,\Phi)-\int_{\partial I_1}{\cal K}_b (\delta_\epsilon\Phi,\Phi),
\end{split}
\ee
for appropriate functions ${\cal K}_0, {\cal K}_b$. 
The above off-shell equations are of course also true on-shell. Then, using the fact that $S=S_0+S_b$ has a well-defined action principle \eqref{variation-principle-def}, we arrive at
\begin{align}\label{chargeIK}
\int_{I_2}\I_0
+{\cal K}_0
+   \int_{\partial I_2} \I_b
    +{\cal K}_b
    \approx
 \int_{I_1}\I_0
 +{\cal K}_0
 +\int_{\partial I_1} \I_b
 +{\cal K}_b,
\end{align}
where $\approx$ denotes on-shell equality. Since $I_1,I_2$ are arbitrary constant time slices, the quantity
\eq{
Q_\epsilon[\Phi]\equiv \int_{I}\I(\delta_\epsilon \Phi, \Phi)+
    \int_{C=\partial I} \Ci(\delta_\epsilon\Phi ,\Phi),
}{charge definition}
with
\be
\I(\delta_\epsilon \Phi, \Phi)\equiv \I_0(\delta_\epsilon \Phi, \Phi)+{\cal K}_0(\delta_\epsilon \Phi, \Phi),\qquad 
\Ci(\delta_\epsilon\Phi ,\Phi)\equiv \I_b(\delta_\epsilon \Phi, \Phi)+{\cal K}_b(\delta_\epsilon \Phi, \Phi),
\ee
defines  a \emph{conserved charge} for solutions of equations of motion, associated with symmetry $\delta_\epsilon$.

\subsection{Electric charges }
Now we are ready to apply the above analysis to the action \eqref{improvedaction}. The integration on the $\tau$-constant hypersurface $I$  denoted as the $\mathcal{I}$-term in \eqref{charge definition} yields\footnote{We note that ${\cal K}_0=0$ in our case, if the source $\mathbf{J}$ is turned off.}, 
\begin{align}
\int_{I}\I
 &=-\frac{1}{p!}\int_{I}d^{d-1}\sigma_{\alpha}\  \partial_{[\mu_0}\Lambda_{\mu_1\cdots \mu_p]} \F^{\alpha\, \mu_0\cdots \mu_p}
 \approx-\frac{1}{p!}\int_{\partial I}\sqrt{\mathscr{G}}\,\tau_b\laa_{a_1\cdots a_p} F^{\rho b\, a_1\cdots a_p}\,,\label{surface-term-1}
    \end{align}
where $d^{d-1}\sigma_{\alpha}$ is the volume form on  $I$, $\tau^b$ is future-directed and  in the last equality we have used the on-shell condition.
The $\mathcal{C}$-term integral in \eqref{charge definition} acquires two contributions on the boundary of $I$; 
\be\label{surface-term-2}
\int_{C}\mathcal{C}
        =-\frac{1}{p!}\int_{C}\sqrt{\mathscr{G}}\tau_b\left((p+1)\partial^{[b}\laa^{a_1\cdots a_p]}{A^\rho}_{a_1\cdots a_p}+2p \laa_{\rho\,a_2\cdots a_p}A^{\rho\,b\, a_2\cdots a_p}\right)\,,
\ee
The first term in \eqref{surface-term-2} comes from the $\B$-term in \eqref{dBterm} ($\I_b$) and the second term in \eqref{surface-term-2} from variation of the boundary term \eqref{Bterm1}  (${\cal K}_b$) and is non-zero for $p>0$.
Putting the $\I$-term  \eqref{surface-term-1} and the $\mathcal{C}$-term \eqref{surface-term-2} into \eqref{charge definition}, the conserved charge for our $(2p+4,p)$-form theory takes the following form
\begin{align}\label{charges1}
Q_{\laa}[\A]=-
\frac{1}{p!}\int_{C}\sqrt{\mathscr{G}}\,\tau_b\,\Big[\left(\laa\cdot F^{\rho b}\right)_p+\left((\di\laa)^b\cdot{A^\rho}\right)_p-2p \left(\laa_{\rho}\cdot A^{\rho b}\right)_{p-1}\Big]\,,
\end{align}
where the integration is over the $C=S^{2p+2}$ at large $\rho$-constant and arbitrary $\tau$-constant hypersurface.
We can write it down  in terms of differential forms  on $S^{2p+2}$; the relevant ones being the $p$-forms $\hat{\mathbf{F}}_{\rho\tau},\, \hat{\mathbf{A}}_\rho,\, \hat{\boldsymbol{\laa}}$ and the $(p-1)$-forms $\hat{\boldsymbol{\laa}}_\rho$ and $\hat{\mathbf{A}}_{\rho\tau}$ (\emph{cf}. notation introduced in section \ref{sse:notation}):
\be\label{charge-expression-form}
Q_{\boldsymbol{\laa}}[\A]=-\int_C\Big(\hat{\boldsymbol{\laa}}\wedge\star\hat{\mathbf{F}}_{\rho\tau}+\hat{\mathbf{A}}_\rho\wedge\star(\widehat{\di\boldsymbol{\laa}})_\tau-2\hat{\boldsymbol{\laa}}_\rho\wedge\star\hat{\mathbf{A}}_{\rho\tau}\Big)\,.
\ee
In the second term, the de Sitter exterior derivative acts on $\boldsymbol{\laa}$ first, then projecting on $S^{d-2}$ yields the sphere $p$-form $(\widehat{\di\boldsymbol{\laa}})_\tau$.

\subsection{Magnetic charges}\label{sec-magnetic}

For a $(p+1)$-form theory in $2p+4$ dimensions, the Hodge star operator maps the $(p+2)$-form field strength $\bcF_{p+2}$ to its Hodge dual which is another $(p+2)$-form, 
\begin{equation}
\bcF_{p+2}\to \star \bcF_{p+2}.
\end{equation}
The source-free equations of motion $\di\star \bcF=0$ which are relevant to the asymptotic region, will allow for a magnetic potential $\star \bcF_{p+2}=\di\boldsymbol{\tilde{\mathcal A}}_{p+1}$. The action is invariant under `magnetic' gauge transformations $\boldsymbol{\tilde{\mathcal A}}\to\boldsymbol{\tilde{\mathcal A}}+\di \boldsymbol{\tilde{\Lambda}}_p$, so we can ask about the conserved charges corresponding to this gauge symmetry. 

The first question is whether the magnetic charges contain new independent information about the fields. We have seen that the gauge potential can be decomposed into de Sitter differential forms:
\begin{equation}\label{dsdecomposition}
\A_{\nu_0\cdots \nu_{p}}:\qquad \left\{
\begin{array}{ll}
A_{\rho\,a_1\cdots a_p}& \ \text{de Sitter $p$-form}\,,\\
A_{a_0\cdots a_p}& \ \text{de Sitter $(p+1)$-form}\,.
\end{array}
\right.
\end{equation}
We showed above that the conserved charges are built out of the de Sitter $p$-form $A_{\rho\,a_1\cdots a_p}$ and its corresponding field strength, independent of the de Sitter $(p+1)$-form $A_{a_0\cdots a_p}$ in \eqref{dsdecomposition}. Similarly, the magnetic charges involve only the $\tilde{A}_{a_1\cdots a_p\rho}$ components which are related to the electric gauge potentials by
\begin{equation}
(\di\mathbf{\tilde A})_{a_0\cdots a_p\rho}=\frac{1}{(p+2)!}{\epsilon^{b_0\cdots b_{p+1}}}_{a_0\cdots a_p\rho}(\extd\bA)_{b_0\cdots b_{p+1}}\,.
\end{equation}
Thus, the magnetic charges extract the information contained in $A_{a_0\cdots a_p}$ being missed by the electric charges.

The expression for the magnetic charges and their conservation follows exactly alongside the discussions we had about electric charges, providing that the same boundary conditions as in  \eqref{bnyc23} are satisfied by the magnetic potentials $\tilde{A}$;
\begin{align}Q_{\tilde{\laa}}[\tilde{\A}]=-
\frac{1}{p!}\int_{C}\sqrt{\mathscr{G}}\,\tau_b\,\Big[\left(\tilde{\laa}\cdot \tilde{F}^{\rho b}\right)_p+\left((\di\tilde{\laa})^b\cdot{\tilde{A}^\rho}\right)_p-2p \left(\tilde{\laa}_{\rho}\cdot \tilde{A}^{\rho b}\right)_{p-1}\Big]\,.
\end{align}

\subsection{Algebra of charges}\label{charge-algebra-sec}

We already presented a general formula for the conserved charge associated with residual gauge transformations \eqref{charge-expression-form}. This formula is linear in gauge transformation parameter and also linear in the background gauge field. The expression for the charge may then be viewed as a functional over the phase space of form-field configurations. One can then compute algebra of charges (Poisson bracket of charges over the phase space).\footnote{In the Hamiltonian or the covariant phase space method for computing charges, as discussed in appendices \ref{Hamiltonian} and \ref{se:covariant},  one computes the charge variation and then integrating over a phase space; whereas we obtain the charge itself in the `Noether's method' proposed above.} The expression for the charge \eqref{charge-expression-form} only involves the $\bA_\rho$ component with the associated gauge transformation
\eqs{
\delta_{\boldsymbol{\epsilon}}\mathbf{A}_\rho=-\di{\boldsymbol\epsilon}_\rho\,.
}
Therefore, the charge algebra is
\eqs{
\{Q_{\boldsymbol{\laa}},Q_{\boldsymbol{\epsilon}}\}=\delta_\epsilon Q_{\boldsymbol{\laa}} =-\int_{C}\Big(\di\boldsymbol{\epsilon}_\rho\wedge\star\di\boldsymbol{\laa}-2\boldsymbol{\laa}_\rho\wedge\star\di\boldsymbol{\epsilon}_\rho\Big)\,.
}
Next, recall that $\boldsymbol{\lambda}$ and $\boldsymbol{\epsilon}$ are $p$-forms generating residual gauge transformations on $dS_{2p+3}$ satisfying \eqref{residual0}, which at leading order takes the following form;
\be\label{lambda-eom}
\di\star\di\boldsymbol{\laa}=2\star\di\boldsymbol{\laa}_\rho\,(-1)^p\,.
\ee
One therefore finds after an integration by parts
\be\label{charge-algebra-general}
\{Q_{\boldsymbol{\laa}},Q_{\boldsymbol{\epsilon}}\}=2\int_{C }\Big(\boldsymbol{\laa}_\rho\wedge\star\di\boldsymbol{\epsilon}_\rho-\boldsymbol{\epsilon}_\rho\wedge\star\di\boldsymbol{\laa}_\rho \Big)
\,.
\ee
In terms of forms on sphere, it becomes
\be\label{charge-algebra-general2}
\{Q_{\boldsymbol{\laa}},Q_{\boldsymbol{\epsilon}}\}=2\int_{C }\Big(\hat{\boldsymbol{\laa}}_\rho\wedge\star(\widehat{\di\boldsymbol{\epsilon}_\rho})_\tau-\hat{\boldsymbol{\epsilon}}_\rho\wedge\star(\widehat{\di\boldsymbol{\laa}_\rho})_\tau\Big)
\,.
\ee
This expression takes a simpler form in temporal gauge $\laa_{\rho\tau B_3\cdots B_p}=0$, 
\be\label{charge-algebra-general3}
\{Q_{\boldsymbol{\laa}},Q_{\boldsymbol{\epsilon}}\}=2\int_{C }\Big(\hat{\boldsymbol{\laa}}_\rho\wedge\star\partial_\tau\hat{\boldsymbol{\epsilon}}_\rho-\hat{\boldsymbol{\epsilon}}_\rho\wedge\star\partial_\tau\hat{\boldsymbol{\laa}}_\rho\Big)
\,.
\ee
The charge algebra can hence be non-Abelian only if the $\rho$-component of the gauge parameters are non-zero. Moreover, the RHS of the charge algebra \eqref{charge-algebra-general}, being independent of the gauge field $\A$, is a $c$-number over the phase space; i.e., the RHS is a central term. In section \ref{se:6dim} we will explicitly compute the charges as well as their algebra in six dimensions for the $(6,1)$-form theory and find the corresponding central charge.

\subsection{Classification of charges}\label{Charge-classification-sec}

We showed that conserved charges of the $(2p+4,p)$-form theory are related to those residual gauge transformations generated by $p$-forms $\boldsymbol{\Lambda}$ on the Minkowski spacetime which preserve our boundary conditions \eqref{bnyc23} and satisfy $\did\di\boldsymbol{\Lambda}=0$ as explained above \eqref{residual0}. To leading order in $\rho$, this equation can be written down in terms of de Sitter forms and operators
\eq{\did\di\blaa+2\di\blaa_\rho=0\,,}{residual dS}
with immediate implication that $\did\di\blaa_\rho=0$.
As mentioned earlier, \eqref{residual dS} specifies $\boldsymbol{\Lambda}$ up to closed forms, i.e. there is a `gauge symmetry' in this equation. We can fix this extra freedom through a `gauge fixing' condition. A convenient choice especially for the computation of charges is the \emph{temporal gauge fixing}\footnote{One could have fixed de Sitter  or  Minkowski covariant gauges, respectively,  $\did\blaa=0$ or $\did\boldsymbol{\Laa}=0$. Our results on charge algebra and classification is of course independent of this gauge fixing on $\blaa$.};
\eq{\boldsymbol{\tau}\cdot\boldsymbol{\Lambda}=0\qquad\text{or}\qquad \hat{\blaa}_\tau=\hat\blaa_{\rho\tau}=0\,.}{temporalgauge}
In this gauge upon decomposing forms on de Sitter in terms of forms on sphere we have $\blaa=\hat{\blaa}$ and we may only work with hatted gauge parameter forms which live on the sphere. In particular --- see appendix \ref{laplacedS},
\begin{align}\label{temporal-lambda}
 \di\boldsymbol{\laa}&=\Big(\di\tau\wedge\frac{\partial}{\partial\tau}\,+\hat{\di}\,\Big)\blaa
= \di\tau\wedge\dot{\hat{\blaa}}+\hdi\hat{\blaa}\,,\\
 \di^\dagger\boldsymbol{\laa}&=\frac{1}{\cosh^2\tau}\,\hdi^\dagger\hat{\blaa}\,.
\end{align}

We may now discuss solutions to the residual gauge condition equation  \eqref{residual dS}. The simplest solution to this equation is $\di\blaa=0$, which yields $\di\blaa_\rho=0$ where in the temporal gauge \eqref{temporalgauge}, results in,
\be\label{zero-mode-lambda}
\hat{\blaa}=\hdi{\hat\beps}\,, \qquad \dot{\hat{\beps}}=0\qquad\text{and}\qquad\hat{\blaa}_\rho=\hdi{\hat\beps}_\rho\,,\qquad \dot{\hat{\beps}}_\rho=0\,.
\ee
This is the \emph{zero-mode} solution. It turns out that in this case $\hat{\blaa}_\rho$ does not contribute to the charge and one may choose $\hat{\beps}_\rho=0$. Other solutions to the equation of residual gauge condition \eqref{residual dS} can be classified by decomposing it on the sphere in the temporal gauge \eqref{temporalgauge};
\begin{subequations}\label{3}
\begin{align}
\hdid\dot{\hat{\blaa}}+2\dot{\hat{\blaa}}_\rho\cosh^{2}\tau&=0\,,\label{31}\\
\hdi^\dagger\hdi\hat\blaa+\partial_\tau\left(\cosh^2\tau\,\dot{\hat{\blaa}}\right)+2\hdi\hat{\blaa}_\rho\cosh^{2}\tau&=0\,.\label{32}
\end{align}
\end{subequations}
where \eqref{31} is the projection of \eqref{residualdS} on the $\tau$ direction.
We note that \eqref{3} is written in terms of $p$ or $(p-1)$ forms on the sphere $S^{2p+2}$ for which we can use Hodge decomposition. Except for the $p=0$ case, the Hodge decomposition (see appendix \ref{Ap-diff}) will not involve the harmonic part.  So for the moment we focus on the $p\neq 0$ case and shall consider the harmonic case later. In order to solve \eqref{3} we then write
\be
\hat\blaa= \hat\blaa^{\text{\tiny exact}}+\hat\blaa^{\text{\tiny coexact}}.
\ee
Eq.\eqref{3} then splits into three equations,
\begin{subequations}\label{344}
\begin{align}
\hdid\dot{\hat{\blaa}}^{\text{\tiny exact}}+2\dot{\hat{\blaa}}_\rho\cosh^{2}\tau&=0\,,\label{4a}\\
\partial_\tau\left(\cosh^2\tau\,\dot{\hat{\blaa}}^{\text{\tiny exact}}\right)+2\hdi\hat{\blaa}_\rho\cosh^{2}\tau&=0\,,\label{4b}\\
\hdi^\dagger\hdi\hat\blaa^{\text{\tiny coexact}}+\partial_\tau\left(\cosh^2\tau\,\dot{\hat{\blaa}}^{\text{\tiny coexact}}\right)&=0\,.\label{4c}
\end{align}
\end{subequations}
The above analysis explicitly shows that equations \eqref{344} have three different classes of solutions for $\hat{\blaa}$. Upon changing variables as $y=\tanh \tau$, these cases read as:
\begin{itemize}
\item \textbf{Coexact.} $\hat{\blaa}=\did{\hat{\boldsymbol{\psi}}}$ is coexact and together with $\hat{\blaa}_\rho$ are solutions to 
\eqref{4c};
\eq{
(1-y^2)\hat{\blaa}^{\prime\prime\text{\tiny coexact}}
+\hat{\Delta}\hat{\blaa}^{\text{\tiny coexact}}=0\,,\quad \hat{\blaa}_\rho\cong0\,,\quad {\hat{\blaa}}^\prime_\rho=0\,.}{5c}
where ${}^\prime$ is derivative w.r.t. the new variable $y$ and $\cong$ denotes  `equality up to an exact/harmonic form'.

\item \textbf{Exact.} $\hat{\blaa}=\di\hat{\beps}$ is an exact form and in general time-dependent, subject to \eqref{4a}-\eqref{4b}. In this case $\hat{\beps}$ is specified in terms of  $\hat{\blaa}_\rho\ncong 0$,
\begin{align}\label{5ab}
\hat{\Delta}\hat{\beps}^{\prime}=\frac{2\hat{\blaa}^\prime_\rho }{(1-y^2)}\,,\quad
\hat{\beps}^{\prime\prime}\cong-\frac{2{\hat{\blaa}}_\rho}{(1-y^2)^2}\,,\quad \did\di{\hat\blaa}_\rho=0
\end{align}
\item \textbf{Zero-mode.} As spelled out in \eqref{zero-mode-lambda}, $\hat{\beps}$ is constant in time, with no restriction on its angular dependence. Zero-modes form a  subclass of exact parameters, identified by setting $\blaa_\rho\cong 0$ in \eqref{5ab}.\footnote{In $(6,1)$-theory, $\beps$ is a function. Actually the time-dependent $\beps(y,\hat{x})=c y$ parameter solves \eqref{5ab} with vanishing, $\blaa_\rho$. However, its charge is always vanishing, as the strings pierce the sphere at two points with opposite contributions.}
\end{itemize}

We have discussed exact and coexact parts of $\hat{\blaa}$ and recalling the Hodge theorem, there remains  $\hat{\blaa}^{\text{\tiny harmonic}}$. The only harmonic forms on an $n$-sphere are the constant function, and the volume form. Top-form gauge parameters are discussed in \cite{Chernyavsky:2017xwm}. In the context of $(2p+4,p)$-form theory the only relevant case is $p=0$ where the charges are point-like and the sources do not reach celestial sphere $S^2$. In this case, however, the gauge parameter is a 0-form and the Hodge decomposition allows for a harmonic part, a constant function on  $S^2$. This constant function gives rise to the zero-mode charge.

One may note that, as the definition above indicates, all the zero-modes are also exact. However, the zero-modes are distinct, as they are exact over the whole de Sitter and Minkowski spacetime and not just on codimension 2 asymptotic sphere. These two have also different  physical meanings. The zero-mode part, as discussed, corresponds to the `global part' of gauge transformations which keep a given gauge field configuration $\bcA$ intact and the corresponding charges may be computed using the standard Noether's theorem (see \cite{Hajian:2015xlp} for more discussions); these charges are the usual brane charges. The zero-mode charges and the coexact charges commute with themselves and with other exact charges, while the exact charges are not commuting; their algebra has a central extension. As our explicit example in \eqref{exact-charge-commutator} shows, the fact that exact charges do not commute, at the technical level, is due to the fact that $\blaa_\rho\ncong0$ for the exact charges.
{From now on we  omit the hat $\,\hat{}\,$ from operators and forms on sphere and only retrieve it when necessary.}

\section{Two concrete examples}\label{sec:charge-computation-examples}

In this section we explicitly compute the charges \eqref{charge-expression-form} for two $p=0,1$ examples. The $p=0$ case has been studied (extensively) in the literature \cite{Strominger:2013lka,Kapec:2015ena,Campiglia:2015qka,Seraj:2016jxi,Strominger:2017zoo} and we show how our formulation recovers/compares with those analysis. The $p=1$, however, is new and has the novel feature that it contains non-Abelian sector of `exact charges'.

\subsection{Maxwell theory in four dimensions (\emph{p=0})}\label{se:4dim}

The $(4,0)$-form theory is defined by the following action,
\begin{equation}\label{(4,0) B-term}
    S=-\frac12\int_{\mathcal M}\sqrt{-g}\F_{\mu\nu}\F^{\mu\nu}-\int_{B}\sqrt{-h}\,A^\rho A_\rho\,,
\end{equation}
and the boundary conditions \eqref{bnyc1}, \eqref{bnyc2} and \eqref{bnyc3} which translate to,
\begin{equation}
\A_{\rho }=\frac{A_{\rho }}{\rho}+\ordr
{-2}\,,\qquad \A_{a}=A_{a}+\ordr
{-1}\,.\label{(4,0) gauge field b.c.}
\end{equation}
In this case the gauge parameter is a $p=0$ form, i.e. a scalar, and \eqref{(4,0) gauge field b.c.} is preserved by large gauge transformations, if the gauge parameter has the boundary behavior,
\eq{\Lambda=\lambda+{\mathcal O}(\rho^{-1})\,.}{BPGT4,0} 
It follows that gauge transformations at leading order in $\rho$ are
\eqs{
\delta_{{\laa}} A_\rho=0\,,\qquad \delta_{{\laa}} A_a=\partial_a\laa\,.
}
The improved action \eqref{(4,0) B-term} is  gauge invariant under the boundary condition preserving gauge transformations \eqref{BPGT4,0} and together with the boundary conditions \eqref{(4,0) gauge field b.c.} defines a well-posed action principle in the Lorenz gauge,
$\nabla_\mu \A^{\mu}=0$.
To leading order in $\rho$, equations of motion and the Lorenz gauge condition are
\begin{subequations}
\begin{align}
    D_aD^aA_\rho&=0\,,\label{eom401}\\
    D^aD_{[a}A_{b]}&=0\,,\label{eom402}\\
    D^aA_a+2A_\rho&=0\,.\label{lgauge40}
\end{align}
\end{subequations}
In this case we do not have a $\lambda_\rho$ part and \eqref{residual dS} reduces to 
\eq{D_aD^a\lambda=0\,.}{residual40}

As discussed in section \ref{section-3}, surface charges \eqref{charge definition} consist of two terms. The first term leads to a surface integral on the boundary of $I$, lying on  $\rho=\rho_0$ and $\tau=\tau_0$ surface;
\begin{equation}\label{Iterm40}
\int_{I}\I(\delta_{{\laa}} \A,\A)=
\int_{I}\partial_\mu\Lambda\, \F^{\mu\nu}d^3\sigma_\nu\approx\int_{C}\sqrt{\mathscr{G}}\tau_a\laa\, \partial^a A_\rho \,,
\end{equation}
where in the second equality we have used the on-shell condition $\partial_\mu{\cal F}^{\mu\nu}=0$ and the boundary conditions \eqref{(4,0) gauge field b.c.}-\eqref{BPGT4,0}. 
The second surface term in the charge \eqref{charge definition} comes from the $\B$-term \eqref{dBterm}, the remainder after subtraction of the varied boundary term \eqref{Bterm0};
\eqs{\int_{C}\Ci(\delta_{\laa} \A ,\A)
=-\int_{C}\sqrt{\mathscr{G}}\tau_a\partial^a\laa A_\rho \,.
}
Putting these together (that is writing \eqref{charge-expression-form} for $p=0$ case), we recover the same expression of \cite{Campiglia:2017mua} for the surface charge
\eqs{\label{(4,0) charge}
Q_{\blaa}[\A]=\int_{S^2}\sqrt{\mathscr{G}}(\laa \partial^a A_\rho-A_\rho \partial^a\laa)\tau_a\,,
}
where $\tau_a$ is the future directed timelike normal vector of $C=S^2$. Since the action \eqref{(4,0) B-term} is strictly gauge invariant within boundary conditions \eqref{(4,0) gauge field b.c.}, the charges are conserved once equations of motion hold. The conserved charge \eqref{(4,0) charge} is a functional of $A_\rho(\tau, \hat{x}^a)$, which is a function on sphere at any given $\tau$. 

\subsubsection{ Solution space and the antipodal matching}\label{antipodal-matching-4d}
 The charge has been written in terms of two scalars $A_\rho$ and $\laa$ on the de Sitter spacetime $dS_3$,  both satisfying the wave equation \eqref{eom401} and \eqref{residual40}.
General solution to this equation is found by expanding in spherical harmonics
\begin{equation}\label{gensol40}
A_\rho(\tau,\hat{x})=\sum_{lm}Y_{lm}(\hat{x})f_l(y)\,,\qquad -l\leq m\leq l\;,\; l\geq0\,.
\end{equation}
where $y=\tanh \tau$ and $f_l(y)$ satisfies
\begin{equation}
(1-y^2)f_l^{\prime\prime}(y)+l(l+1)f_l(y)=0\,.\label{tau diff}
\end{equation}
There are two independent solutions to this equation;
\eq{f_l(y)=c_1f_l^{(1)}(y)+c_2\,f_l^{(2)}(y)\,.}{}
The zero mode solution is;
\eq{ f_0^{(1)}(y)=1\,,\qquad f_0^{(2)}(y)=y\,,
}{}
which for $q=c_2$ this describes the famous electric monopole solution with the electric charge $q$ --- see the footnote \ref{zeromode40}. 

For all $l\geq1$ we have, 
\begin{align}
f^{(1)}_l(y)&=
(1-y^2)^{\frac{1}{2}}P^1_l(y)\,
,\qquad
f^{(2)}_l(y)=
(1-y^2)^{\frac{1}{2}}Q^1_l(y)\,,
\label{tau solution}
\end{align}
where $P^1_l$ and $Q^1_l$ are Legendre functions of the first and the second kind respectively. To confirm the conservation, we notice that because of the integration on $S^2$, the charge \eqref{(4,0) charge} picks up the modes with the same $l$ and opposite $m$. In addition, its $\tau$-dependence is the Wronskian of the differential equation \eqref{tau diff} which is a constant. So the charge is independent of $\tau$ (or $y$) and is \emph{conserved}.

\paragraph{Antipodal matching.} The fields appearing in charge expressions, those in \eqref{gensol40}, are defined on the global $dS$ space $B$. It is, however, known that not all points on a global $dS$ space are causally connected (see  Fig. \ref{dS-Penrose}). 
In particular, one can distinguish two regions which are separated by the cosmological horizon and the physical fields/observables may be defined only on one side of this horizon. One way to define fields on the whole $dS$ is to define them on one cosmological patch and extend it to the other side by the `antipodal matching' \cite{Strominger:2001pn}. In our conventions the point $(\tau, \hat{x}^A)$ is mapped onto $(-\tau,-\hat{x}^A)$ by antipodal map (see Fig. \ref{dS-Penrose}). In particular, in our case we require 
 \begin{equation}
 F_{a\rho}(-\tau,-\hat{x})=F_{a\rho}(\tau,\hat{x}),
 \end{equation}
where $F$ is the gauge field strength on $dS_3$. 
Requiring the above antipodal matching, we learn that 
 \begin{equation}
 A_\rho(-\tau,-\hat{x})=-A_\rho(\tau,\hat{x})\,.
 \end{equation}
 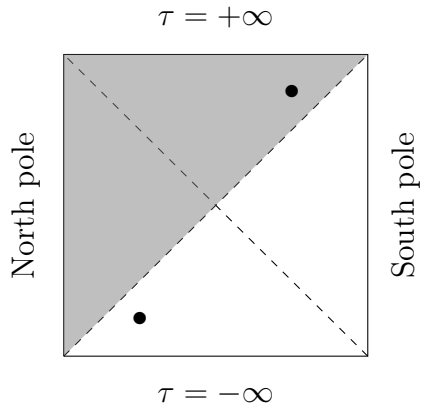
\begin{figure}
     \centering
\begin{tikzpicture}
\draw (0,0)--(0,4)--(4,4)--(4,0)--(0,0);
\draw[dashed](0,0)--(4,4);
\draw[dashed](4,0)--(0,4);
\node [rotate=90] at (-.5,2) {North pole};
\node [rotate=90] at (4.5,2) {South pole};
\path[fill=gray,opacity=0.5](0,0)--(0,4)--(4,4)--(0,0);
\node [] at (2,-.5) {$\tau=-\infty$};
\node [] at (2,4.5) {$\tau=+\infty$};
\node [] at (3,3.5){$\bullet$};
\node [] at (1,.5){$\bullet$};
\end{tikzpicture}
     \caption{Penrose diagram of de Sitter space. The shaded region is the causal future of the north pole, and its dashed $45^{\circ}$ boundary is the future horizon of the south pole. The antipodal map amounts to  a couple of horizontal and vertical flips. Especially, the north pole at far past is mapped to the south pole at far future.}
     \label{dS-Penrose}
 \end{figure}

The above condition, recalling the parity and time-reversal properties of the modes, 
\begin{equation}\label{parity transform}
P^1_l(-y)=(-1)^{l+1}P^1_l(y)\,,\quad  Q_l^1(-y)=(-1)^{l}Q_l^1(y),\quad
Y^m_l(-\hat{x})=(-1)^lY_l^m(\hat{x})\,,
\end{equation}
leads to the fact that there should not be $Q_l$ modes in the gauge field.
So, the general allowed solution for $A_\rho$ is
\begin{equation}
A_\rho(y,\hat{x})=qy+(1-y^2)^\frac12\sum_{l\geq 1,m} \frac{a_{lm}}{\sqrt{{l(l+1)}}}Y_{l}^m(\hat{x})
P^1_{l}(y)\,.\label{gauge field expansion}
\end{equation}
We comment that the boundary term \eqref{(4,0) B-term} remains finite if $A_\rho$ does not involve $Q_l$ modes, as
\footnote{The zero-mode $l=0$ solution ${\bar A}_\rho=c_1+q\tanh\tau $, corresponds to an electric background gauge field due to a point-like source whose asymptotic $\tau$-behavior is drastically different from higher modes. The boundary term \eqref{(4,0) B-term} as it stands is divergent for this mode. As discussed in section \ref{variational-section} around Eq. \eqref{varbackground} about a a well-defined action principle in such a case should be refined. Explicitly, take the decomposition $\A_{\rho }=\left(\bar{A}_\rho+A_\rho\right)/\rho+\ordr{-2}$ where $\delta\bar{A}_\rho=0$ and $A_\rho$ is a square-integrable function on $dS_3$. The resulting phase space includes all configurations with fixed total charge $q$. In this case $\delta S\sim  \int_{S^2}\delta A_\tau$, for recovering 
the action principle we need to either restrict our boundary conditions s.t. $\int_{S^2}\delta A_\tau$ or improve the action by adding a new boundary term proportional to $\int_{S^2}A_\tau$
\label{zeromode40}.}
$$
\int_{-1}^{1} \di y(1-y^2)^{-1}P^1_{l}(y)P^1_{l^{\prime}}(y)\delta_{ll^\prime}=\frac{(l+1)!}{(l-1)!}\de_{l,l^\prime}\,.
$$

On the other hand, for having non-zero charges, the gauge parameter $\laa$ should have the following form;
 \begin{equation}
\laa(y,\hat{x})=-\frac{\laa_0}{4\pi}+(1-y^2)^{\frac{1}{2}}\sum_{l\geq1,m} \frac{\laa_{lm}}{\sqrt{l(l+1)}}Y_{l}^m(\hat{x})
Q^1_{l}(y)\,,\label{gauge parameter expansion}
\end{equation}
hence for the gauge parameter,
  \begin{equation}
  \laa(-\tau,-\hat{x})=\laa(\tau,\hat{x}),
  \end{equation}
and the reality of the gauge parameter and the gauge field implies  
$$a_{lm}=a^\ast_{l,-m}, \qquad \laa_{lm}=\laa^\ast_{l,-m}.$$  
\paragraph{Expression of the charge.} For generic gauge field $A_\rho(y,\hat{x})$ and gauge parameter $\laa(y,\hat{x})$ the conserved charge becomes
\eqs{\label{(4,0) charge evaluated}
Q_\laa[\A]=\int_{S^2}d^2\hat{x}(\partial_y\laa
A_\rho-\partial_yA_\rho\laa)=q\laa_0+ \sum_{l\geq1,m} a_{lm}\laa^\ast_{lm}\,.}

Defining $\laa_+(\hat{x})\equiv \laa(\tau\to+\infty,x^A) $ and $\laa_-(\hat{x})\equiv\laa(\tau\to-\infty,\hat{x})$ and denoting the charges computed at $\tau\to\pm$ by $Q^\pm$ , we recover the antipodal matching condition for the charges proved in \cite{Campiglia:2017mua}
\begin{equation}\label{antipodal-charges}
Q^{+}_{\laa_+}[\A]=Q^{-}_{\laa_-}[\A].
\end{equation}
As discussed above, the antipodal matching and finiteness of the boundary term and asymptotic charges in our analysis yield to the same condition (absence of $Q_l$ modes in the gauge field). Moreover, the antipodal matching in our setup is a physically well-motivated requirement as it is the natural way to extend definition of physical field on global $dS$. This may be compared with other arguments for antipodal matching \cite{Strominger:2013lka, Campiglia:2017mua}. It is desirable to explore and understand the antipodal matching  better.

The $l=0$ term in \eqref{(4,0) charge evaluated} is the contribution of the zero-mode charge corresponding to a global transformation on the sphere, $\Delta{\laa}=0$, (\emph{cf.} section \ref{zero-mode}). We denote $l\geq 1$ terms as coexact charges as they correspond to coexact gauge transformations on $S^2$, $\di^\dagger{\laa}=0$. There are no exact charges  in $4d$ as we are dealing with a scalar gauge parameter $\lambda$. Moreover, the zero-mode and usual coexact charges, as \eqref{(4,0) charge evaluated} suggests, can be recombined into charges associated with a coclosed $\lambda$ on the $S^2$. This is in fact the more usual viewpoint used to describe multipole charges of Maxwell theory \cite{He:2014cra, Seraj:2016jxi, Campiglia:2017dpg}. 

Computation of magnetic charges would lead to the same result \eqref{(4,0) charge evaluated} with the integrand replaced by magnetic potential $\tilde{A}$ and the corresponding gauge parameter $\tilde{\laa}$. The magnetic zero-mode charge would give the total number of magnetic monopoles.

\subsection{$2$-form theory in six dimensions (\emph{p=1})}\label{se:6dim}

In this section we study the simplest yet non-trivial case with charges associated to exact gauge transformation on $S^4$, the $p=1$ case. The improved action is
\begin{equation}\label{(6,1) improved action}
    S=-\frac{1}{12}\int_{\mathcal M}\sqrt{-g}\F_{\mu\nu\alpha}\F^{\mu\nu\alpha}-\int_{B}\sqrt{-h}\,A^{\rho a} A_{\rho a}\,.
\end{equation}
with the boundary conditions,
\begin{equation}
\A_{ \rho a}=\frac{A_{\rho a}}{\rho}+\ordr{-2}\,,\qquad \A_{ab}=A_{ab}+\ordr{-1}\,.\label{(1,6) gauge field b.c.}
\end{equation}
To leading order in $\rho$, the field equations and the Lorenz gauge condition are,
\begin{subequations}
\begin{align}
D^bF_{\rho\,ba}=0\,,\qquad  &D^aA_{\rho\,a}=0\,,
\label{radial-eom}\\
D_cF^{cab}=0\,, \qquad &D_b A^{ba}+2A^{\rho\,a}=0\,.\label{transvers eom}
\end{align}\label{eom16}
\end{subequations}
 To leading order in $\rho$, the $\rho$-component of the field strength $ \F_{\rho\, a b}= \partial_\rho \A_{ab}+\partial_b \A_{\rho a}+\partial_a \A_{b\,\rho}$,
 yields;
\begin{equation}\label{fabr16}
F_{ab\,\rho}=2\partial_{[a}A_{b]\,\rho}\,,
\end{equation} 
which means  $F_{ab\,\rho}$ is a closed 2-form on $dS_5$ with the potential $A_{\rho\,a}$. 
\subsubsection{Exact, coexact and zero-mode conserved charges}\label{6dcharges}

The parameters of the boundary condition preserving gauge transformations
 at leading order in $\rho$
generate the following transformations on the boundary 
\eqs{
\delta A_{a\rho}=\partial_a\laa_\rho\,,\qquad \delta A_{ab}=2\partial_{[a}\laa_{b]}\,.
}
They include  a de Sitter scalar $\lambda_\rho$ and a de Sitter vector $\laa_a$ which satisfy \eqref{residual dS} in temporal gauge, i.e.
\begin{subequations}\label{Lambda eq}
\begin{align}
    &D^aD_{[a}\laa_{b]}=D_b\laa_\rho\,, \label{residual16}\\
    &\laa_\tau=0\,.\label{residual163}
\end{align}
\end{subequations}
As discussed in section \ref{Charge-classification-sec}, the whole contribution of the gauge parameters to the expression of charges, is reduced to knowing $\hat{\lambda}_B$ on $S^4$. The solutions to these equations come in three classes: zero-mode solutions for which $\boldsymbol{\lambda}=\extd\boldsymbol{\epsilon}$ on $dS_{5}$, and the exact (coexact) charges for which $\hat{\boldsymbol{\lambda}}$ is an exact (coexact) 1-form on $S^4$.
Below we discuss each cases separately.
\paragraph{Coexact charges.}
In this case $\laa_B=\D^C\psi_B{}_C$ while, $\laa_\rho=$ const. and $\laa_\tau=0$ as a consequence of \eqref{Lambda eq}. Consequently, $\delta_\laa A_{a\rho}=0$ and hence the improved action \eqref{(6,1) improved action} is strictly gauge invariant, so the situation is exactly similar to the Maxwell in four dimensions i.e. the $(4,0)$ theory. There remains two contributions to the expression of the charge  \eqref{charges1}  with gauge parameters being coexact on $S^4$. For constant $\tau$ slices and in the temporal gauge, the expression for coexact charges  \eqref{(6,1) coexact charge} simplifies to
\begin{align}\label{(6,1) coexact charge}
Q^{\text{\tiny coexact}}_{\blaa}[\A]
=-\int_{S^4}\sqrt{\mathscr{G}}\Big(\partial_\tau\laa_{B}A^{\rho B\,} -\laa^B\, \partial_\tau  {A^\rho}_{B}\Big),
\end{align}
where in the first term above we used  \eqref{fabr16}. The second term in \eqref{(6,1) coexact charge} is the contribution of $\delta_\lambda S_0$ denoted as the $\B$-term in \eqref{dBterm}.
We also notice that in \eqref{(6,1) coexact charge}  both $\hat\blaa=\laa_B \extd x^B$ and $\bA_\rho=A_{\rho B}\extd x^B$ are coexact 1-forms on $S^4$ as a consequence of orthogonality of exact and coexact forms.
Finally, since $\delta_\lambda A_{\rho B}=0$ in this case (as $\laa_\rho=$ const.), these charges commute among themselves;
\begin{align}
   \{ Q^{\text{\tiny coexact}}_\epsilon,Q^{\text{\tiny coexact}}_\laa\}=-\delta_\eps Q^{(\text{\tiny coexact})}_\laa=0\,.
\end{align}

\paragraph{Exact charges.} These charges are associated with gauge transformations generated by the de Sitter scalar $\laa_\rho$ and the de Sitter vector $\lambda_a$ which are related via solving \eqref{Lambda eq}.
Expression for the charges \eqref{charges1} with the gauge parameters being exact on $S^4$ takes the form
\begin{align}\label{(6,1) exact charge}
Q^{\text{\tiny exact}}_{\blaa}[\A]
=-\int_{S^4}\sqrt{\mathscr{G}}\,\Big( \partial_\tau\laa_{B}A^{\rho B\,}-2\laa_\rho A^{\rho \tau}\Big)\,,
\end{align}
where the last term in the expression above is coming from variation of the boundary term $\delta_\lambda S_b$ in \eqref{Bterm1}.
Since $\laa_B$ is an exact form on $S^4$, one can readily see that the coexact part of the gauge field $A_{B\rho}$ which is divergence-free, drops from the expression of the charge and only the exact part of the background gauge field contribute to \eqref{(6,1) exact charge}. This already proves that the commutator of exact charges with the coexact ones are zero,
\begin{align}
   \{ Q^{\text{\tiny coexact}}_\epsilon,Q^{\text{\tiny exact}}_\laa\}=0\,.
\end{align}
We may now define  $\laa_B=\partial_B\epsilon$ and $A^{\text{\tiny exact}}_{\rho B}=\partial_B\phi$ for some functions $\epsilon$ and $\phi$. It can be shown  that on-shell --- see appendix \ref{Appendix-forms}, $A_{\rho\tau}=\partial_\tau\phi$; so \eqref{(6,1) exact charge} simplifies to
\begin{align}\label{(6,1) exact charge1}
 Q^{\text{\tiny exact}}_{\blaa}[\A]&=-\int_{S^4}\sqrt{\mathscr{G}}\,\Big(2\laa_\rho \partial_\tau{\phi}+\partial_\tau\partial_B\epsilon D^B\phi\Big)\nn\\
 &=2\int_{S^4}\sqrt{\mathscr{G}}\,\Big(\phi\partial_\tau\laa_\rho-\laa_\rho \partial_\tau\phi\Big)\,,
\end{align}
where in the last line we made an integration by part and also used \eqref{4b}; $\mathcal{D}^B\mathcal{D}_B\dot{\epsilon}=2\partial_\tau\laa_\rho \cosh^2\tau$. 

The more interesting part is the commutator of exact charges: the central charge for the exact charge sector turns out to be non-zero. The reason is simply that  $\laa_\rho\neq0$  in this case and the gauge transformation on $A^{\text{\tiny exact}}_{\rho B}$ (and $A_{\rho \tau}$) is non zero and acts on $\phi$ as a shift
\eqs{
\phi\to \phi-\laa_\rho.
}
We consequently find
\begin{align}\label{exact-charge-commutator}
\{
Q^{\text{\tiny exact}}_\epsilon,Q^{\text{\tiny exact}}_\laa\}
=-\delta_\eps Q^{(\text{\tiny exact})}_\laa
=2\int_{S^4}\sqrt{\mathscr{G}}\Big[-\epsilon_\rho\partial_\tau\laa_\rho+\laa_\rho \partial_\tau\epsilon_\rho
\Big]\,.
\end{align}

\paragraph{Zero-mode charges.}
We  discussed above exact gauge parameters with $\laa_\rho\neq 0$. It remains to consider exact gauge parameters with $\laa_\rho=0$. For this class, $(\di\laa)_{BC}=0$ on the sphere, so they leave the gauge parameter invariant; they are \emph{exact symmetries} \cite{Hajian:2015xlp} and the charge reduces to (\emph{cf.} section \ref{zero-mode})
\be\label{zero-mode-6d}Q^{\text{\tiny zero-mode}}_\laa[\A]=\int_{S^4}\sqrt{\mathscr{G}}\,\laa_BF^{\rho \tau B}.
\ee
These charges are non-vanishing only in the presence of sources that pierce the celestial sphere, like infinite strings. Zero-mode charges obviously commute with all charges in the theory.

\subsubsection{Solution space and mode expansion}\label{solution space 6,1}
To compute the explicit expression of the charges we need to solve the equations  for gauge potentials and gauge parameters. 
Field equations \eqref{eom16} and \eqref{Lambda eq} describe all components of the gauge field and the gauge parameter respectively. We notice that since $A_{\rho a}$ is a $1$-form on the $dS_5$ background, equations \eqref{radial-eom} are essentially Maxwell's equations in Lorenz gauge for the 1-form gauge field $A_{\rho a}$ and take the following form
\begin{subequations}
\begin{align}
(1-y^2)A^\pp_{\rho\tau}+4yA^\p_{\rho\tau}+\Big[\Delta+4\frac{1+y^2}{1-y^2}\Big]A_{\rho\tau}&=0\,,\label{Aeta eq}\\
(1-y^2)A^\pp_{\rho B}+\Delta A_{\rho B}+\frac{2y}{1-y^2}\partial_BA_{\rho\tau}&=0\,,\label{A_B eq}\\
A_{\rho\tau}^\p-\mathcal{D}^BA_{\rho B}+\frac{4y}{1-y^2}A_{\rho\tau}&=0\,.\label{Lorenz eq}
\end{align}
\end{subequations}
where prime is derivative w.r.t. $y=\tanh\tau$ and $\Delta$ is the Laplace-Beltrami operator on the 4-sphere. 
The spectrum of $\Delta$ acting on functions and $1$-forms is given in the appendix \ref{Ap-diff}. Solutions to \eqref{Aeta eq}  are
\eqs{
A_{\rho \tau}=(1-y^2)^{\frac{3}{2}}
\sum_{l\geq0,m_\alpha}\Big[
c^{(1)}_{lm_\alpha}P^1_{l+1}(y)+c^{(2)}_{lm_\alpha}Q^1_{l+1}(y)\Big]Y_{lm_\alpha}(\hat{x})\,,
\label{Aeta solution}
}
where $Y_{lm_\alpha}(\hat{x}), \alpha=1,2,3$ are spherical harmonics on the $4$-sphere, e.g. see \cite{Higuchi:1986wu,Akhoon:2014lda}.
If $A_{\rho B}$ is an exact form, then $A_{\rho B}(x^a)=\partial_B\phi(x^a)\equiv A_{\rho B}^{\text{\tiny exact}}$  and the Lorenz condition \eqref{Lorenz eq} leads to
\eqs{\label{lorenz with phi}
\mathcal{D}^2\phi=A^\p_{\rho \tau}+\frac{4y}{1-y^2}A_{\rho \tau}\,.
}
Plugging \eqref{Aeta solution} into  \eqref{lorenz with phi}, and expanding in eigen-modes  $\mathcal{D}^2\phi=\sum_{l\geq0} l(l+3)\phi_l$ as explained in the appendix \ref{Ap-diff}, after some manipulations one gets,
\eqs{\label{6,1 A exact solution}
A_{\rho B}^{\text{\tiny exact}}=(1-y^2)
\sum_{l>0,m_\alpha}\frac{1}{l(l+3)}\Big[c^{(1)}_{lm_\alpha}P^2_{l+1}(y)+c^{(2)}_{lm_\alpha}Q^2_{l+1}(y)\Big]\partial_B Y_{lm_\alpha}(\hat{x})\,,
}
where $c^{(1)}_{lm_\alpha}$ and  $c^{(2)}_{lm_\alpha}$ are the same as in \eqref{Aeta solution} and $A_{\rho B}^{\text{\tiny exact}}=0$ for $l=0$. 
In the appendix \ref{Appcoexact}, we also verify that up to the zero mode on sphere, $A_{\rho \tau}=\partial_\tau \phi$ where $\phi$ is a solution to $D^aD_a\phi=0$. Thus $A_{\rho a}=\partial_a\phi\equiv A_{\rho a}^{\text{\tiny exact}}$ is a \emph{pure gauge} de Sitter vector with $F_{\rho\, ab}=0$ which could be eliminated from the beginning using residual gauge transformations.\footnote{The operator $\mathcal{D}^2$ annihilates $l=0$ mode of $\phi$, so it is not invertible. However, since we are looking for  $A^{\text{\tiny exact}}_{\rho B}=\partial_B\phi$, that mode is unconstrained by $A^{\text{\tiny exact}}_{\rho B}$. One is free to choose it such that $A_{\rho \tau}\big|_{l=0}=\partial_\tau\phi\big|_{l=0}$, as well as higher-$l$ modes.} As we have already  seen in section  \eqref{6dcharges}, these pure gauge configurations can possess non-zero exact charges.

The field equation \eqref{A_B eq} has an exact and a coexact part which are linearly independent and should be individually zero. For $A_{\rho B}$ being coexact denoted as $A_{\rho B}^{\text{\tiny coexact}}$, on the sphere, it  simplifies as
\eqs{
(1-y^2)A^\pp{}^{\text{\tiny coexact}}_{\rho B}+\Delta A^{\text{\tiny coexact}}_{\rho B}=0\,,\label{non-exact eq}
}
with solutions
\eqs{
A^{\text{\tiny coexact}}_{\rho B}(y,\hat{x})=(1-y^2)^{\frac{1}{2}}\sum_{lm_\alpha}\omega_B^{lm_\alpha}(\hat{x})\Big[b^{(1)}_{lm_\alpha}P^1_{l+1}(y)+ b^{(2)}_{lm_\alpha}Q^1_{l+1}(y)
\Big]\,\label{coexact solution}
,}
where $ \omega_B^{lm_\alpha}$ are coexact eigen 1-forms of the Laplace-Beltrami operator on $S^4$;
\eq{\Delta_H\omega^{lm_\alpha}_B=[l(l+3)+2]\omega_B^{lm_\alpha}\qquad\text{with}\qquad l\geq 1\,.}{}

\paragraph{Solving the gauge parameter.} 
Among all equations in \eqref{Lambda eq} for the gauge parameter, only the following equations are independent,
\begin{subequations}\label{laa-6d}
\begin{align}
(1-y^2)\laa^{\pp\text{\tiny coexact}}_B+\Delta \laa^{\text{\tiny coexact}}_B
&=0\label{laa eq2}\,,\\
(1-y^2)\laa^\pp_\rho+2y\laa^\p_\rho+\Delta\laa_\rho&=0\,,
\label{laa eq4}
\end{align}
\end{subequations}
while other components of exact/coexact gauge parameters are specified from \eqref{5c}-\eqref{5ab}. The general solution for $\laa^{\text{\tiny coexact}}_B$ is the same as in \eqref{coexact solution} and will be discussed in the next section. The general solution for $\laa_\rho$ is
\eq{\laa_\rho(y,\hat{x})=\lambda^{(1)}_0+\lambda^{(2)}_0(\tfrac{y^3}{3}-y)+ (1-y^2)\sum_{l>0, m_\alpha}\big[\lambda^{(1)}_{lm_\alpha} {\cal P}_{lm_\alpha}(y,\hat{x})+ \lambda^{(2)}_{lm_\alpha}{\cal Q}_{lm_\alpha}(y,\hat{x})\big],}{laa rho-solution}
where ${\cal P}, {\cal Q}$ are the following functions,
\begin{subequations}\label{PQfunctions}
\begin{align}
{\cal P}_{lm_\alpha}(y,\hat{x})&\equiv\sqrt{\frac{2(l-1)!}{(l+3)!}}Y_{lm_\alpha}(\hat{x})P^2_{l+1}(y)\,,\\
{\cal Q}_{lm_\alpha}(y,\hat{x})&\equiv
\sqrt{\frac{2(l-1)!}{(l+3)!}}Y_{lm_\alpha}(\hat{x})Q^2_{l+1}(y)\label{normalized-modes}\,, 
\end{align}
\end{subequations}
with $l\geq 1$. These functions on $dS_5$ are normalized as:
\eq{
\int \di y\di^4\hat{x}\sqrt{-h}(1-y^2)^2 {\cal P}_{lm_\alpha}(y,\hat{x}) {\cal P}_{l'm'_\alpha}(y,\hat{x})=\de_{ll}\delta_{m_\alpha,m'_\alpha},
}{normalization-PQ}
and similarly for ${\cal Q}$'s.

\subsubsection{Evaluation of charges and the central extension}

We may now compute the explicit expression for the coexact and exact charges \eqref{(6,1) coexact charge} and \eqref{(6,1) exact charge1} in terms of the mode expansions of the gauge parameters and gauge fields contributing to these charges.

In the coexact sector, both the gauge parameter $\laa_B$ and the gauge field $A^{\text{\tiny coexact}}_{B\rho}$ satisfy same equations \eqref{laa eq2} and \eqref{non-exact eq} with a general solution \eqref{coexact solution}. As in the coexact charges of the Maxwell theory in four dimensions i.e. $(4,0)$-theory, discussed in section \ref{antipodal-matching-4d}, not all terms in  \eqref{coexact solution} keep our boundary term \eqref{Bterm0} finite. 
It turns out that similar to our argument in the $(4,0)$-theory, $Q^m_l$ solutions in \eqref{PQfunctions} are not square-integrable and make our boundary term \eqref{Bterm0} divergent. 
Moreover, the finite contribution to the charge comes from the $Q^1_{l+1}(y)$ term in $\lambda_B$. That is, 
\begin{align}
    A^{\text{\tiny coexact}}_{B\rho}(y,\hat{x})&=(1-y^2)^{\frac{1}{2}}\sum_{lm_\alpha}\omega_B^{lm_\alpha}(\hat{x})\Big[\frac{b_{lm_\alpha}}{\sqrt{(l+1)(l+2)}}P^1_{l+1}(y)\Big]\,,\\
    \laa^{\text{\tiny coexact}}_B(y,\hat{x})&=(1-y^2)^{\frac{1}{2}}\sum_{lm_\alpha}\omega_B^{lm_\alpha}(\hat{x})\Big[\frac{\laa_{lm_\alpha}}{\sqrt{(l+1)(l+2)}}Q^1_{l+1}(y)\Big]\,,
    \label{laaB solution}
\end{align}
 for $l\geq 1$. Now the expression of the charge in \eqref{(6,1) coexact charge} is the Wronskian of \eqref{non-exact eq} and by orthonormality of $\omega_B^{lm_\alpha}$ it yields
\eqs{
Q^{\text{\tiny coexact}}_\laa[\A]=\sum_{l\geq 1, m_\alpha} b_{lm_\alpha}\laa^\ast_{lm_\alpha}\,.
}

The explicit expression for the exact charges \eqref{(6,1) exact charge1} can be given in terms of the mode expansions for $\laa_\rho$ given in \eqref{laa rho-solution} and those for $A^{\text{\tiny exact}}_{\rho B}=\partial_B\phi$ in \eqref{6,1 A exact solution} where both $\laa_\rho$ and $\phi$ satisfy the same equation as in \eqref{laa eq4}. 
The crucial point is that in contrast to the coexact part,  
as shown in the appendix \ref{Appcoexact}, for exact gauge fields the boundary term \eqref{Bterm0} is a total derivative, thus we need not disallow one of the branches allowed by equations of motion.\footnote{The boundary term $S_b$ in this case describes a massless scalar on $dS_{5}$ which can be regularized by holographic renormalization means \cite{deHaro:2000vlm}.}
On the other hand, unlike the coexact case, both $\lambda^{(1)}_{lm_\alpha}$ and $\lambda^{(2)}_{lm_\alpha}$ modes in \eqref{laa rho-solution} can contribute to exact charges and we hence have two sets of exact charges, which will conveniently be denoted by $Q^{(a)}_{lm_\alpha}, a=1,2$.  In this respect, the exact charges are different than the coexact and zero-mode charges. However, one class of the parameters $\lambda^{(a)}_{lm_\alpha}$, say $a=1$ leads to non-zero charges only if $\phi$ belongs to the opposite class, which has opposite behavior under PT. 
 As a result, the exact charges exhibit antipodal matching property too as will be discussed below.

The more interesting part is, however, the algebra of the exact charge sector, for which we need to evaluate \eqref{exact-charge-commutator}, which by using
the general solution \eqref{laa rho-solution} for $\laa_\rho$ and $\epsilon_\rho$, and recalling the normalized Legendre functions \eqref{normalized-modes} and \eqref{normalization-PQ}, yields
\be\label{charge-algebra}
\{
Q^{(a)}_{lm_\alpha},Q^{(b)}_{ l^\prime m^\prime_\alpha}\}=4\de _{ll^\prime}\de _{m_\alpha,-m_\alpha^\prime}
\epsilon^{ab},\qquad l\geq 1, \quad a,b=1,2,
\ee
where $\epsilon^{ab}$ is the anti-symmetric symbol.\footnote{As mentioned $\lambda^{(1)}_{lm_\alpha}, \lambda^{(2)}_{lm_\alpha}$ behave oppositely under parity. Nonetheless, the expression of commutator of exact charges \eqref{exact-charge-commutator} involves a time derivative and hence the expression receive a non-zero contribution for $Q^{(1)}_{lm_\alpha}, Q^{(2)}_{lm_\alpha}$ charges. }

\paragraph{Antipodal matching.} As mentioned in \ref{antipodal-matching-4d} the causal connection between points on $\I^+$ and $\I^-$ as boundaries of the de Sitter slices is made via null geodesics beginning on the sphere at $\I^-$ and reaching its antipode at $\I^+$ \cite{Strominger:2001pn}. This would verify the antipodal matching property of the field strength for generic $(2p+4,p)$-theories as we explicitly did for the $(4,0)$-theory in section \ref{antipodal-matching-4d},
\eqs{
\hat{\mathbf{F}}_{\rho\tau}(-\tau,-\hat{x})=
\hat{\mathbf{F}}_{\rho\tau}(\tau,\hat{x})\,.\label{antipodality}} 
 This condition restricts the $p$-form $\hat{\mathbf{A}}^{\text{\tiny coexact}}_\rho$ to its one branch of solutions similar to \eqref{coexact solution} and satisfying \eqref{non-exact eq} on $S^{2p+2}$. In particular, the time-dependence will be given  by $P^1_{l+p}(y)$ while the spacial-dependence on $S^{2p+2}$ is governed by $\omega^{lm_\alpha}_{(p)}(\hat{x})$ with the following parity transformations \cite{Folland1989}
\begin{align}
\label{paritytransformp}
P^1_{l+p}(-y)=(-1)^{l+p+1}P^1_{l+p}(y)\,,\quad
\omega^{lm_\alpha}_{(p)}(-\hat{x})=(-1)^{l+p}\omega^{lm_\alpha}_{(p)}(\hat{x}),\quad l\geq1\,.
\end{align}
Thus, $\hat{\mathbf{A}}^{\text{\tiny coexact}}_\rho$ is odd under PT; implying that the field strength $\hat{\mathbf{F}}_{\rho\tau}$ is even as in \eqref{antipodality}. Similarly one can confirm that the gauge parameters entering the coexact charges are also even. In the exact sector, the field strength is identically zero so both solutions are allowed. In the case of exact charges, antipodal matching is a consequence of finiteness of charges. One may verify that the zero-mode charges also satisfy antipodal matching conditions discussed above. Hence all charges of the ($2p+4,p$)-theory in general satisfy a relation like \eqref{antipodal-charges}, explicitly, 
\eqs{
Q^{+}_{\boldsymbol{\laa}_+}[\mathbf{A}]=Q^{-}_{\boldsymbol{\laa}_-}[\mathbf{A}]\,.
}

\subsubsection{Summary of the asymptotic charges in 6d $2$-form theory}

We discussed that there are three classes of charges for the 2-form theory in six dimensions i.e. $(6,1)$-form theory. The zero-mode charges are specified by time-independent $0$-forms (the $\epsilon$) on $S^4$. These charges may be denoted by 
$Q^{\text{\tiny zero-mode}}_{l m_\alpha},\ l\geq 0$. The exact and coexact charges are respectively specified by exact and coexact 1-forms on the $S^4$. There is one set of coexact charges $Q^{\text{\tiny coexact}}_{l m_\alpha},\ l\geq 1$ but two sets of exact charges $Q^{(a)}_{lm_\alpha},\ l\geq 1, a=1,2$. The zero-mode and coexact charges commute with all other charges and only the exact charges of different kind do not commute. Their commutator is given in \eqref{charge-algebra} which is an infinite set of Heisenberg algebras. In deriving these algebras \eqref{charge-algebra} we assumed $\lambda$ is independent of the gauge field $\bA$. One may construct other algebras through quadratic combination of these charges associated to linearly field-dependent gauge parameters \cite{Afshar:2016wfy,Afshar:2016kjj}. 
This point needs further analysis which we hope to perform in future works.

The coexact and zero-mode charges have correspondents in the usual $4d$ Maxwell theory, but the exact charges are new objects. They have a feature that they are conserved even off-shell, since the field strength is zero in their case. This feature is reminiscent to the case of asymptotic charges associated to Weyl transformation in conformal gravity \cite{Afshar:2011yh,Afshar:2011qw,Afshar:2013bla} where the value of the Weyl factor is in no way restricted by field equations.
As a comment on the physical meaning of these exact charges, we note that the same expression as in \eqref{(6,1) exact charge} appears for electric conserved charges in the 6d Maxwell theory i.e. the $(6,0)$-form theory with the same boundary conditions given in \eqref{bnyc1}; see \cite{Ortaggio:2014ipa} for further analysis.  The asymptotic de Sitter space has topology $\mathbb{R}_\tau\times S^4$ with trivial first cohomology group. As a result, for a  purely electric configuration, that is $F_{ab}=0$, one has   $A_a=\partial_a\phi$ for some de Sitter scalar $\phi$. Applying the same procedure for conserved charges as we did for the $(4,0)$-form theory (with Lorenz gauge replaced by radial gauge $A_\rho=0$), one obtains \eqref{(6,1) exact charge} with $A_{\rho a}=-\partial_a\phi$ which is exact.


\section{Discussion and outlook}\label{discussion-section}

In this work we studied asymptotic symmetries of $(p+1)$-form theories in `critical' $2p+4$ dimensions. This specific dimension has the remarkable feature that the radiation and Coulomb fields have the same fall-off behavior \cite{Ortaggio:2014ipa, Campoleoni:2017qot}. Although it is expected to have memory effect for general $(d,p)$-form theory, this feature  brings the possibility of having more interesting  $p$-form memory effect in the critical dimensions. The $p$-form memory effect, compared to the usual gravitational or electromagnetic cases in four dimensions, has the novel feature that objects carrying the $(p+1)$-form charges are $p$-branes which have internal degrees of freedom and have the possibility of altering their shape as a $(p+1)$-form photons pass by. Our analysis here has set the stage for studying such $p$-form memory effects.

We showed that the systematic treatment of surface charges leads to their classification into \emph{coexact, zero-mode} and \emph{exact} charges. This classification is of course  Lorentz invariant. As $(d,p)$-form theories are generalizations of electrodynamics ($p=0$), the zero-mode and coexact charges have electromagnetic analogues and have Abelian charge algebras. The exact charges, however, appear in $p\geq 1$ and have non-Abelian algebra. We presented explicit computation of exact charge algebra for $(6,1)$-form theory which we found to be an infinite set of Heisenberg algebras; similar result is expected for generic $(2p+4,p)$-form theories. It is desirable to better understand this algebra and its potential physical observable effects. 

The zero-mode charges were shown to be relevant to the first law of black $p$-brane thermodynamics \cite{Compere:2007vx}. Furthermore, it was shown in  \cite{Copsey:2005se} that for black holes with non-trivial horizon topology, dipole charges can also contribute to the first law. One may then ask whether the class of zero-mode charges can also contribute to the first law of thermodynamics for black branes of non-trivial horizon topology. 

In this work, we computed asymptotic charges at spatial infinity and worked in Lorenz gauge. Moreover, to impose the fall-off behavior we used de Sitter slicing of flat space (\emph{cf.} Fig.\ref{Mink-de-Stter-comparison}), as was done \emph{e.g.} in \cite{Campiglia:2015qka} and \cite{Mann:2008ay}. One may wonder how much the final results on soft charge algebra depend on the slicing and the gauge condition. Although appropriate choice of slicing facilitates imposing fall-off behavior and the boundary conditions, it should not alter the final result once we fix the boundary conditions. For example, to analyze the problem with the boundary condition which is usually set to capture radiation (null rays) reaching the infinity, it is more appropriate to use null slicing.  As another example, for the  Hamiltonian approaches  to asymptotic symmetries at spatial infinity, however, it is more appropriate to use the standard $(t,r)$-slicing, in which all constant time $t$ slices are mapped to a same constant  time $\tau$ surface at large $\rho$ in the de Sitter slicing. Therefore, comparing the asymptotic charges and their algebras for these three cases should be handled with care. This point is pertinent for the case of 3d or 4d gravity where we are dealing with $\mathfrak{bms}_3$ and $\mathfrak{bms}_4$ algebras, and has been noted and analyzed in \cite{Troessaert:2017jcm, Compere:2017knf, Henneaux:2018cst} and for 4d Maxwell theory in \cite{Campiglia:2015qka, Campiglia:2015lxa, Campiglia:2017mua}.

In principle, the soft charges and their algebra are expected to be gauge independent for a given theory defined by boundary fall-off conditions and a given boundary term. However, we should note that to have a well-defined theory, besides an action and boundary conditions, one needs  a suitable boundary term. The latter is needed to ensure having a well-defined variation principle. As we discussed, the form of this boundary term, besides boundary fall-off dependence, also does depend on the gauge-fixing condition (albeit only its asymptotic form) and it affects the expression for the soft charges.

In our derivation of asymptotic charges we first fixed the boundary term needed for having a well-defined action principle. This kind of analysis is well established  in the context of AdS/CFT and holographic renormalization. Among other things, we showed that finiteness of this boundary term is consistent with antipodal matching of asymptotic charges. One may explore other physical effects our boundary value problem may bear, possibly  to establish a flat-space holography for these $(d,p)$-form gauge theories. In particular, the bulk $p$-form gauge symmetry  in these $(d,p)$-form theories has a residual part after gauge fixing; appearing as a boundary $(p-1)$-form gauge symmetry on asymptotic $dS$ slices. This latter may be viewed as `holographic' dual of the bulk theory. Another interesting question in this line is how the arguments here for $p$-form residual symmetries could be combined with the notion of (higher-form) generalized global symmetries \cite{Gaiotto:2014kfa,Grozdanov:2016tdf,Grozdanov:2017kyl,Hofman:2017vwr}.

Among the class of $(d,p)$-form theories with $d=2p+4$, the odd $p$ cases, especially $p=1,3$ cases are of great interest. For odd $p$,  self-dual form field theories with real forms are possible. The self-dual two-form and four-form cases appear in the context of six and ten dimensional supergravity theories. In these cases the equations of motion are first order and there exists a different class of background solutions. For these cases we expect a mixing between the electric and magnetic asymptotic charges we discussed here. It is desirable to explore this case in more detail.

Finally, we point out that although we worked with scale invariant quantities throughout the paper, the charges are conserved up to $1/\rho$ corrections. In this sense, the charges are \emph{asymptotic}, defined at $\rho\to\infty$. An interesting question is how they could be defined inside the bulk as symplectic symmetries \cite{Sheikh-Jabbari:2016lzm} of the theory.  The asymptotic charges can gather more information from the fields in the interior of the spacetime, if divergent $\rho ^n,\, n>0$ gauge parameters are taken into account. These are called \emph{multi-pole charges} \cite{Seraj:2016jxi,Compere:2017wrj} due to their relation to multi-pole moments of fields and the sources.

\acknowledgments

We would like to thank Miguel Campiglia, Kamal Hajian, Vahid Hosseinzadeh and Ali Seraj for fruitful discussions and Andrea Campoleoni, Geoffrey Compere, Dario Francia, Daniel Grumiller, Carlo Heissenberg, Marc Henneaux and Andy Strominger for comments on the draft. H.A. was partially supported by Iran's National Elites Foundation (INEF). The work of M.M.Sh-J and E.E. has been supported in part by the junior research chair in black hole physics of Iranian NSF, project no 951024. They also acknowledge the ICTP NT-04 network scheme.

\appendix

\section{Differential forms on sphere}\label{Appendix-forms}

\subsection{Hodge decomposition}\label{Ap-diff}
We will state some definitions and propositions about differential forms on a sphere \cite{Warner1983,Folland1989}. Given a compact oriented Riemannian $n$-manifold $\M$, the Hodge star operator $\star$ maps any $p$-form to  a $(n-p)$-form, and satisfies
\eqs{\label{hodgesquare}
\star\star=(-1)^{p(n-p)}.
}
The co-differential operator $\di^\dagger$ acting on $p$-forms is defined by\footnote{Both \eqref{hodgesquare} and \eqref{codif} acquire one more minus sign when the signature is Lorentzian.},
\eqs{\label{codif}
\di^\dagger=(-1)^{n(p+1)+1}\star\di\star}
 and acts as
 \eqs{
 (\di^\dagger\omega)_{B_2\cdots B_p}=-\mathcal{D}^{B_1}\omega_{B_1\cdots B_p}\,.
 }
 The Laplace-Beltrami operator $\Delta$ is defined by
\eqs{
\Delta=\di\di^\dagger+\di^\dagger\di\,,
} and if $\Delta\alpha=0$, then $\alpha$ is called a \emph{harmonic} form. A differential form is harmonic  iff
\be
\di\alpha=\di^\dagger\alpha=0\,.
\ee
One can define an inner product on the space of $p$-forms:
\eqs{
\langle\alpha,\beta\rangle=\int_\M\alpha\wedge\star\beta\,.\label{inner product}
}
It follows that $\di^\dagger$ is the adjoint of $\di$
\eqs{
\langle\di\alpha,\beta\rangle=\langle\alpha,\di^\dagger\beta\rangle\,,
}
while $\Delta$ is self-adjoint. 

According to the Hodge decomposition theorem, any differential $p$-form  on a closed Riemannian manifold $\M$ can be decomposed into exact, coexact and harmonic forms:
\eqs{
\omega_p=\di\alpha_{p-1}+\extd^\dagger\beta_{p+1}+\gamma_p\,.
}
The number of harmonic $p$-forms on $\M$ is equal to the dimension of the $p$-th de Rham cohomology group $\mathbf{H}_{\text{dR}}^p(\M)$. On an $n$-sphere, de Rham cohomology is trivial unless $p\in \{0,n\}$ where it becomes isomorphic to $\mathbb{R}$. Since we are mainly interested on $p$- or $(p+1)$-forms on a $(2p+2)$-sphere, no harmonic forms are present in our discussion for $p\geq1$.
Eigenvalues of Laplace-Beltrami operator acting on $p$-forms on an $n$-sphere differ for exact and coexact forms:
\begin{align}
\Delta\omega_{pl}^{\text{\tiny exact}}&=\left[l(l+n-1)+(p-1)(n-p)\right]\omega_{pl}^{\text{\tiny exact}}\\
\Delta\omega_{pl}^{\text{\tiny coexact}}&=\left[l(l+n-1)+p(n-p-1)\right]\omega_{pl}^{\text{\tiny coexact}}\,.
\end{align}
Exact and coexact forms are orthogonal in the sense of inner product \eqref{inner product},
\eqs{
\langle \di \alpha,\di^\dagger\beta\rangle=\langle\alpha,\di^\dagger\di^\dagger\beta\rangle=0\,.
}
Given a coordinate system $x^A$ on a sphere, exact and coexact $p$-forms satisfy
\eqs{
\di \alpha=0 \leftrightarrow \mathcal{D}_{[B_0}\alpha_{B_1\cdots B_p]}=0\,,\qquad
\di^\dagger \alpha=0 \leftrightarrow \mathcal{D}_{B_1}\alpha^{B_1\cdots B_p}=0\,.
}
Finally, for a function $f$ and a 1-form $\omega=\omega_Bdx^B$ on an $n$-sphere we have
\begin{equation}
\begin{array}{rll}
     \Delta_H f&=- \mathcal{D}^C\mathcal{D}_Cf\,,  \\
     \Delta_H \omega&= -\mathcal{D}^C\mathcal{D}_C\omega+(n-1)\omega\,.
\end{array}
\end{equation}
where $\mathcal{D}$ is the covariant derivative on the $n$-sphere.

\subsection{The Laplace operator on  $S^{n}$ induced from $dS_{n+1}$}\label{laplacedS}
Consider the splitting of general $k$-forms on $dS_{n+1}$ as\footnote{Here we distinguish the quantities on the sphere from the ones on de Sitter space by a hat.}
\eqs{
\boldsymbol{\omega}=\hat{\boldsymbol{\omega}}+\di\tau\wedge\hat{\boldsymbol{\omega}}_\tau\,.
}
The Hodge star and co-differential operators have the following form in terms of those on sphere
\begin{subequations}
\begin{align}
\star\boldsymbol{\omega}&=\Big[(-1)^k\di\tau\wedge\hat{\star}\hat{\boldsymbol{\omega}}-\hat{\star}\hat{\boldsymbol{\omega}}_\tau\cosh^2\tau\Big]\cosh^{n-2k}\tau\,,\\
\cosh^{2}\tau\,\di^\dagger\boldsymbol{\omega}&=\cosh^{2k-n}\tau\partial_\tau\Big(\hat{\boldsymbol{\omega}}_\tau \cosh^{n-2k+2}\tau\Big)+\hat{\di^\dagger}\hat{\boldsymbol{\omega}}-\di\tau\wedge\hat{\di^\dagger}\hat{\boldsymbol{\omega}}_\tau\,.
\end{align}
\end{subequations}
We can now compute the Laplace operator $\Delta$,
\begin{subequations}
\begin{align}
\cosh^2\tau \widehat{\Delta\boldsymbol{\omega}}&=\hat{\Delta}\hat{\boldsymbol{\omega}}+\cosh^2\tau\,\ddot{\hat{\bome}}+\frac{1}{2}(n-2k)\sinh2\tau\dot{\hat{\bome}}
+\sinh2\tau\hat{\di}\hat{\bome}_\tau\,,
\\
\cosh^2\tau \widehat{\Delta\bome}_\tau&=\hat{\Delta}\hat{\bome}_\tau+\cosh^2\tau\,\ddot{\hat{\bome}}_\tau+(n-2k+2)(\hat{\bome}_\tau+\frac12\sinh2\tau \dot{\hat{\bome}}_\tau)\\
&\qquad\qquad\qquad\qquad\qquad\qquad\qquad-2\tanh\tau\,\hat{\di^\dagger}\hat{\bome}\,.\nn
\end{align}
\end{subequations}
Now we apply these relations to our current problem with $n=2p+2\,,k=p$:
\begin{subequations}\label{APPdecompose}
\begin{align}
\cosh^2\tau \widehat{\Delta\boldsymbol{\laa}}&=\hat{\Delta}\hat{\boldsymbol{\laa}}+\partial_\tau\Big(\cosh^{2}\tau\dot{\hat{\boldsymbol{\laa}}}\Big)+2\sinh\tau\cosh\tau\,\hat{\di}\hat{\blaa}_\tau\,,\label{APPdecompose1}
\\
\cosh^2\tau (\widehat{\Delta\blaa})_\tau&=\hat{\Delta}\hat{\blaa}_\tau+\cosh^2\tau\,\ddot{\hat{\blaa}}_\tau
+4\cosh^{2}\tau\partial_\tau(\hat{\blaa}_\tau\tanh\tau)
-2\tanh\tau\,\hat{\di^\dagger}\hat{\blaa}\,\label{APPdecompose2}.
\end{align}
\end{subequations}

\subsection{Exact and coexact  parts of gauge fields/parameters}\label{Appcoexact}

The coordinate $\rho$ is manifestly Lorentz invariant. Consequently, the Lorentz generators $L_{\mu\nu}=x_\mu\partial_\nu-x_\nu\partial_\mu$ can be written in terms of $x^a$ and $\partial_a$ and they turn out to be the isometries of $dS_{d-1}$ which is expected because both represent $\mathfrak{so}(d-1,1)$ algebra. Thus, a Lorentz transformation  is equivalent to a de Sitter coordinate transformation $x^a\to x^{\prime a}$.
These considerations enable us to decompose Minkowski tensors like $\A_\mu$ into  de Sitter representations $\A_\rho$ and $\A_a$. The former is a Lorentz/de Sitter scalar,
while $\A_a$ is a de Sitter vector.

In $(2p+4,p)$-form theories that we are studying, the gauge field is a  $(p+1)$-form $\A_{\mu_0\cdots\mu_p}$ in Minkowski space. The leading terms in asymptotic $\rho$-expansion can be decomposed in a Lorentz-covariant way as
\eqs{
\begin{array}{ll}\label{A-classify1}
    A_{\rho\, a_1\cdots a_p} &\qquad \text{de Sitter $p$-form}, \\
     A_{a_0\cdots a_p} &\qquad \text{de Sitter $(p+1)$-form.}
\end{array}
}
In this paper we are mostly interested in \emph{electric description} of the theory, in which the first row of \eqref{A-classify1} plays the main role, and only those components contribute to all charges. The components in the second row are related to magnetic  charges, so from now on we focus only on $A_{\rho a_1\cdots a_p}$. These components completely determine the $\rho$-component of the field strength tensor
\eqs{
F_{\rho\, a_0\cdots a_p}=-(p+1)\partial_{[a_0}A_{|\rho|\,a_1\cdots a_p]}\,.\label{ap-coex-Field}
}
\paragraph{Field strength and coexact charges.}
First we consider the on-shell field strength and corresponding gauge fields. Taking all indices of \eqref{ap-coex-Field} on sphere, the equation can be written as
\eqs{
\hat{\mathbf{F}}_\rho=-\di\hat{\mathbf{A}}_\rho,\label{ap-co-Bianchi}
}
and  if we split the latter into
 \eqs{
\hat{\mathbf{A}}=
\hat{\mathbf{A}}^{\text{{\tiny coexact}}}_{\rho}+
\hat{\mathbf{A}}^{\text{{\tiny exact}}}_{\rho}\,,
}
clearly the second term is irrelevant to the field strength, thus $\hat{\mathbf{F}}_\rho$ is determined by $\hat{\mathbf{A}}^{\text{{\tiny coexact}}}_{\rho}$.

On the other hand, if one of the indices of $\mathbf{F}_\rho$ is temporal, then
\eqs{\label{F decomposition}
\hat {\mathbf{F}}_{\rho\tau}=-\partial_\tau \hat {\mathbf{A}}^{\text{\tiny coexact}}_{\rho}-\Big(\partial_\tau\hat{\mathbf{A}}^{\text{\tiny exact}}_{\rho}-\di\hat {\mathbf{A}}_{\rho\tau}\Big)\,,
}
and the equation of motion is
\eqs{\label{ap-co-eom-rhotau}
\mathcal{D}^{B_1}{\hat F}_{\rho\tau B_1\cdots B_p
}=0\qquad\text{or}\qquad \di^\dagger\hat{\mathbf{F}}_{\rho\tau}=0.
}
\eqref{ap-co-eom-rhotau} shows that $\hat{\mathbf{F}}_{\tau\rho}$ is a co-closed $p$-form on $S^{2p+2}$, which in turn  implies that the terms in parenthesis must be harmonic forms, and hence vanishing (if $p>0$).
In conclusion, $\hat{\mathbf{F}}_{\tau\rho}$ is also determined solely by $\hat{\mathbf{A}}^{\text{\tiny coexact}}$:
\begin{align}
\hat{\mathbf{F}}_{\rho}&=-\di\hat{\mathbf{A}}^{\text{\tiny coexact}}_\rho\,,\label{ap-co-F-rho}\\
\hat{\mathbf{F}}_{\rho\tau}&=-\partial_\tau\hat{\mathbf{A}}^{\text{\tiny coexact}}_\rho\,.\label{ap-co-F-taurho}
\end{align}
\eqref{ap-co-F-rho} resulted from Bianchi identity, while \eqref{ap-co-F-taurho} was a consequence of field equations, with no use of gauge conditions. We have shown that $F_{\rho\,a_0\cdots a_p}$ is built out of the coexact part of the gauge field $A_{\rho B_1\cdots B_p}$ and this is a Lorentz invariant statement as the indices on $F$ imply. For $p=0$, $\mathbf{A}_\rho$ is a scalar with no exact parts.

\paragraph{Gauge potential and exact charges.}
Let's study the exact part of the gauge field. We argued that the parenthesis in \eqref{F decomposition} vanishes on-shell, thus
\eqs{
\partial_\tau\hat{\mathbf{A}}^{\text{\tiny exact}}_\rho=\di \hat{\mathbf{A}}_{\rho\tau}.
\label{ap-co-13}}
The action \eqref{improvedaction} and in particular its boundary term were derived in the Lorenz gauge \eqref{Lorentz gauge split} and \eqref{LG-2}; the latter being
\eqs{
\di^\dagger \hat{\mathbf{A}}_{\rho\tau}=0.
}
By exactness we introduce $\hat{\mathbf{A}}^{\text{\tiny exact}}_\rho\equiv\di\hat{\boldsymbol{\phi}}$
where the RHS is a $(p-1)$-form on $S^{2p+2}$. From \eqref{ap-co-13} we obtain\footnote{The case of $p=1$ and $\mathbf{A}_{\tau\rho}$  a constant function on $S^4$ must be dealt with separately.}
\eqs{
\di \Big(\hat{\mathbf{A}}_{\rho\tau}-\partial_\tau\hat{\boldsymbol{\phi}}_\rho\Big)=0\quad\Rightarrow\quad \hat{\mathbf{A}}_{\rho\tau}\cong\partial_\tau\hat{\boldsymbol{\phi}}_\rho.
}

In conclusion, the exact and temporal components of gauge field are on-shell built out of a $(p-1)$-form $\hat{\boldsymbol{\phi}}_\rho$ on $S^{2p+2}$ according to
\begin{align}
\mathbf{A}^{\text{\tiny exact}}_\rho&=\di\hat{\boldsymbol{\phi}},\label{ap-coexact-rho}\\
\hat{\mathbf{A}}_{\rho\tau}&\cong\partial_\tau\hat{\boldsymbol{\phi}}.\label{ap-coexact-taurho}
\end{align}
Note the similarity with the system of equations \eqref{ap-co-F-rho},\eqref{ap-co-F-taurho}. Under residual gauge transformations in temporal gauge
\eqs{\delta_{\blaa} \hat{\mathbf{A}}_{\rho}=-\di\hat{\blaa}_\rho,\qquad \delta_{\blaa} \hat{\mathbf{A}}_{\rho\tau}=-\partial_\tau\hat{\blaa}_\rho
\label{ap-co-residual}
}
$\hat{\boldsymbol{\phi}}$ transforms by a shift
\eqs{
\hat{\boldsymbol{\phi}}\to \hat{\boldsymbol{\phi}}-\hat{\blaa}_\rho.
}


\section{Canonical analysis of the $(d,p)$-form theory}\label{Hamiltonian}
This appendix contains in part a review of what appeared in \cite{Henneaux:1992ig, Henneaux:1986ht}. We will denote 
the $d$ dimensional spacetime coordinates by $x^\mu$, the time direction by $x^0$ and its spatial part by $x^i\equiv\ex$. 
The Lagrangian for the Abelian $p$-form gauge theory is
\begin{align}
L&=-\frac{1}{2}\frac{1}{(p+2)!}\int  \di^{d-1}\ex \,\Big[(p+2)\F_{0 i_0\cdots i_{p}}\F^{0i_0\cdots i_{p}}+\F_{i_0\cdots i_{p+1}}\F^{i_0\cdots i_{p+1}}\Big]\nn\\
&=\frac{1}{2(p+1)!}\int  \di^{d-1}\ex \,\Big[(\dot{\A}_{i_0\cdots i_{p}})^2-2(p+1)\dot{\A}_{i_0\cdots i_{p}}\partial_{[i_0} \A_{|0|i_1\cdots i_{p}]}\nn\\
&\qquad\qquad\qquad\qquad\quad
+(p+1)^2(\partial_{[i_0} \A_{|0|i_1\cdots i_{p}]})^2-\frac{1}{(p+2)}\F_{i_0\cdots i_{p+1}}\F^{i_0\cdots i_{p+1}}\Big]\,.
\end{align}

The boundary conditions in spherical coordinates are deduced from Coulomb behaviour of the fields (\emph{cf.} section \ref{se:boundary conditions}),
\eqs{\label{canon. b.c. A}
\A^{0\, B_1\cdots B_p}=A^{0\, B_1\cdots B_p}\,r^{3-d}+\mathcal{O}(r^{2-d}),\qquad \A^{B_0\cdots B_p}=A^{ B_0\cdots B_p}\,r^{2-d}+\mathcal{O}(r^{1-d})\,,
}
where we have imposed radial gauge condition $\A_{rB_1\cdots B_p}=0$. Consequently,  the Coulomb boundary conditions for field strength become \cite{Ortaggio:2014ipa}
\begin{subequations}\label{canon. b.c. F}
\eqs{\F^{0r B_1\cdots B_p}=F^{0r B_1\cdots B_p}\,r^{2-d}+\mathcal{O}(r^{1-d})\,,}
\eqs{\F^{0B_0\cdots B_p}=F^{0 B_0\cdots B_p}\,r^{1-d}+\mathcal{O}(r^{-d})\,,}
\eqs{\F^{rB_0\cdots B_p}=F^{r B_0\cdots B_p}\,r^{-d}+\mathcal{O}(r^{-1-d})\,.}
\end{subequations}

The canonical momenta are defined as
\begin{subequations}
\begin{align}\label{canmumenta1}
\pi^{0i_1\cdots i_p}&\equiv\frac{\partial {\mathcal L}}{\partial \dot{\A}_{0 i_1\cdots i_p}}=0\,,\\
\pi^{i_0\cdots i_p}&\equiv\frac{\partial {\mathcal L}}{\partial \dot{\A}_{i_0\cdots i_p}}=\dot{\A}_{i_0\cdots i_p}-(p+1)\partial_{[i_0}\A_{|0|i_1\cdots i_p]}\,.\label{canmumenta2}
\end{align}
\end{subequations}
where $\mathcal L$ is the Lagrangian density.
Equation \eqref{canmumenta1} just shows that $A_{0 i_1\cdots i_p}$ is not a dynamical field since there is no term in the Lagrangian with time derivative of $A_{0 i_1\cdots i_p}$. Vanishing of the associated momenta constitute the following primary constraints,
\eqs{
\phi_1^{i_1\cdots i_p}\equiv \pi^{0i_1\cdots i_p}=0\,.
}

The canonical commutation relations among the fields on the phase space are
\eqs{
\{A_{\mu_0\cdots \mu_p}(\ex),\pi^{\nu_0\cdots \nu_p}(\text{y})\}=\delta^{d-1}(\ex-\text{y})
\delta_{\mu_0\cdots \mu_p}^{\nu_0\cdots \nu_p}\,,
}
where the generalized Kronecker delta is equal to $+1$ (respectively $-1$) if the lower indices are even (respectively odd) permutations of upper indices, and zero otherwise. 

The canonical Hamiltonian is
\begin{align}
H_C&=\int\di^{d-1}\ex\,\pi^{i_1\cdots i_p}\dot{\A}_{i_1\cdots i_p}-L\nn\\
&=\frac{1}{2}\int\di^{d-1}\ex\Big[
\frac{1}{(p+1)!}\pi_{i_0\cdots i_p}\pi^{i_0\cdots i_p}+
\frac{1}{(p+2)!}\F_{i_0\cdots i_{p+1}}\F^{i_0\cdots i_{p+1}}\,,\nn\\
&\qquad\qquad\qquad-\frac{1}{p!}A_{0i_1\cdots i_p}\partial_k\pi^{ki_1\cdots i_p}
\Big]
+ \mathcal{B}\,.
\end{align}
where the boundary term $\mc{B}$
\eqs{
\mc{B}=\frac{1}{(p+1)!}\int \di^{d-1}\ex \partial_k\big(
\A_{0i_1\cdots i_p }\pi^{ki_1\cdots i_p}
\big)\,,
}
vanishes within boundary conditions \eqref{canon. b.c. A},\eqref{canon. b.c. F}.
There are also secondary constraints
\eqs{
\{\pi^{0i_1\cdots i_p},H_C\}=\partial_k\pi^{ki_1\cdots i_p}\equiv 
\phi_2^{i_1\cdots i_p}\,.
}
The complete set of $p$-form constraints $\phi_1^{i_1\cdots i_p}$ and $\phi^{i_1\cdots i_p}_2$ are first-class and the generator of gauge symmetry will be built out of them. There are no further constraints since $\{\phi_2^{i_1\cdots i_p},H_C\}=0\,$.
\subsection{Gauge transformations, gauge fixing and its reducibility}
The generators of gauge transformation are constructed using the procedure of Castellani \cite{Castellani:1981us};
\eq{
G[\eps]=\frac{1}{p!}\eps^1_{i_1\cdots i_p}\phi_1^{i_1\cdots i_p}
+\frac{1}{p!}\eps^2_{i_1\cdots i_p}\phi_2^{i_1\cdots i_p}
\,.}{gentrans}
where gauge parameters $\epsilon^1$ and $\epsilon^2$ are arbitrary anti-symmetric tensors. The extended action;
\eqs{
S_E=\int \di ^d\ex \Big[
\pi^{i_0\cdots i_p}\dot{\A}_{i_0\cdots i_p}+\pi^{0i_1\cdots i_p}\dot{\A}_{0i_i\cdots i_p}-H_C-u^1_{i_1\cdots i_p}\phi_1^{i_1\cdots i_p}
-u^2_{i_1\cdots i_p}\phi_2^{i_1\cdots i_p}
\Big]\,,
}
is invariant under transformations generated by \eqref{gentrans}. Here $u^1$ and $u^2$ are Lagrange multipliers. 
The gauge transformations $\de_\epsilon F=\{F,G[\epsilon]\}$ read
\begin{align}
&\de \A_{0i_1\cdots i_p}=\eps^1_{i_1\cdots i_p}, &
\de \A_{i_0\cdots i_p}=(p+1)\partial_{[i_0}\eps^2_{i_1\cdots i_p]}\,, \\&\de \pi^{0i_1\cdots i_p}=\pi^{0i_1\cdots i_p}=0\,.
\end{align}
The Lagrange multipliers must transform accordingly to retain invariance of the action \cite{Henneaux:1992ig}:
\eqs{\label{lagrange mult. transformation}
\delta u^1_{i_1\cdots u_p}=\dot{\eps}^1_{i_1\cdots i_p},\qquad \delta u^2_{i_1\cdots u_p}=\dot{\eps}^2_{i_1\cdots i_p}-\eps^1_{i_1\cdots i_p}\,.
}
One usually fixes the Lagrange multipliers corresponding to the secondary and higher generation constraints to zero, reverting to the total action $S_T$ which includes primary constraints only. So we may set $\de u^2=\dot{\eps}^2-\eps^1=0$ in the above,
\begin{align}
&\de \A_{0i_1\cdots i_p}=\dot{\eps}_{i_1\cdots i_p}, &
\de \A_{i_0\cdots i_p}=(p+1)\partial_{[i_0}\eps_{i_1\cdots i_p]}.\label{canonical gauge transf.}
\end{align}

Gauge symmetry of the theory, $A\to A+\di\eps$ involves arbitrary $p$-form gauge parameter $\eps_{\mu_1\cdots\mu_p}$. However this generating set is reducible, since any gauge parameter of the form $\eps=\di\eta$ leaves the fields intact \cite{Henneaux:1986ht, Barnich:2001jy} (in \cite{Hajian:2015xlp} these were called exact symmetries). The reducibility manifests itself in identities among secondary constraints $\phi_2$:
\eqs{
I_{(1)}^{i_2\cdots i_p}=\partial_m \phi_2^{mi_2\cdots i_p}=0.\label{red identity 1}
}
The identities $I_{(1)}$ are not independent either. Taking repetitive divergences produces a chain of identities:
\eqs{
I_{(n-1)}^{i_n\cdots i_p}=\partial_m I_{(n-2)}^{mi_n\cdots i_p},\qquad n=3,\cdots ,p+1\,.
}

Yet another way of spotting the reducibility is through the fact that Noether identities corresponding to the gauge symmetries are not independent,
\eqs{
\nabla_\alpha\nabla_\beta \F^{\alpha\beta \mu_1\cdots\mu_p}=0\,.
}
These are not independent identities, which is evident by taking further derivatives. We can identify the redundant gauge parameters among $\eps_{i_1\cdots i_p}$ in \eqref{canonical gauge transf.}. If \eqs{\eps_{i_1\cdots i_p}=p\partial_{[i_1}\eta_{i_2\cdots i_p]}
\label{eta defined}
}
then the corresponding gauge transformation vanishes
\eqs{
\eps_{i_1\cdots i_p}(\text{x})\{K(\text{y}),\phi_2^{i_1\cdots i_p}(\text{x})\}=0
}
for all $K$ due to reducibility identity $I_1$ \eqref{red identity 1}. However, Lagrange multipliers are left invariant under \eqref{lagrange mult. transformation} only if $\eps_{i_1\cdots i_p}$ is time independent.  The usual procedure in dealing with reducible symmetries, however, is to enhance the gauge symmetry by a set of parameters $\eps_{0i_2\cdots i_p}$ acting only on Lagrange multipliers as
\eqs{
\de u^1_{i_1\cdots i_p}=-p\partial_{[i_1}\eps_{|0|i_2\cdots i_p]}\,.
}
Now the redundant transformations are characterized by $\eps_{0i_2\cdots i_p}=\dot{\eta}_{i_2\cdots i_p}$ where $\eta$ is defined in \eqref{eta defined} and is \emph{arbitrary}. With these considerations, the full set of gauge parameters of the theory contain arbitrary $p$-forms $\eps_{\mu_1\cdots \mu_p}$ in spacetime.

The number of conjugate pairs is $\binom{d}{p+1}$. There are $\binom{d-1}{p}$ primary and $\binom{d-1}{p}$ secondary constraint. Each generation of reducibility identities consists of $\# I_{(n)}=\binom{d-1}{p-n}$ number of relations which should be enumerated by alternating signs. Using Pascal's identity $\binom{d}{p+1}=\binom{d-1}{p+1}+\binom{d-1}{p}$ we can write the whole number as an alternating sum:
\eqs{
\#\text{degrees of freedom}=\sum_{k=0}^{p+1}\binom{d-1}{k}(-1)^{p-k+1}=\binom{d-2}{p+1}\,.
}
One can readily check that  Maxwell's theory $(p=0)$ in $d$ dimensions has $d-2$ degrees of freedom.

\subsection{Surface charges}
The generator of gauge symmetry is explicitly
\eqs{
G[\eps]=\int \di^{d-1}\ex\Big[
\dot{\eps}_{i_1\cdots i_p}\pi^{0i_1\cdots i_p}-\eps_{i_1\cdots i_p}\partial_k\pi^{ki_1\cdots i_p}
\Big]
}

In order to find a differentiable generator, one has to add a boundary term
\eqs{
\de\tilde{G}[\eps]=\de G[\eps]+\de Q[\eps],\qquad \de Q[\eps]=\int\di^{d-1}\ex \partial_k\Big(\eps_{i_1\cdots i_p}\de\pi^{ki_1\cdots i_p}\Big)
}
After this modification, $\tilde{G}[\eps]$ is no longer vanishing on the constraint surface. The on-shell value of $\tilde{G}[\eps]\approx Q[\eps]$ is the surface charge corresponding to transformation generated by $\eps$:
\eqs{
Q[\eps]=\int\di^{d-2}\hat{\ex} \eps_{B_1\cdots B_p}\F^{0rB_1\cdots B_p}\label{hamilton charge}\,.
}
Boundary conditions \eqref{canon. b.c. A} imply that the gauge parameter $\eps$ is $\mathcal{O}(1)$ and it is time-independent. In consequence, the charge \eqref{hamilton charge} is finite. Moreover,  field equations at leading order give
\eqs{
\frac{d}{dt}F_{trB_1\cdots B_p}=0\,,
}
which ensures conservation of the charge.


\section{Covariant approach to charges and their algebra}\label{se:covariant}
We will analyze the gauge symmetry of the system by constructing its covariant phase space, following \cite{Campiglia:2017mua}.

\subsection{Symplectic form and its conservation}
The action may be written more compactly as
\eqs{
S=-\frac{1}{2}\int  \bcF\wedge \ast \bcF
}
the Lee-Wald \cite{Lee:1990nz} symplectic current is a $(d-1)$-form given by
\eqs{
\boldsymbol{\omega}=-\delta_1 \bcA\wedge\delta_2 \ast \bcF-1\leftrightarrow 2\label{symp. current}
}
If the symplectic current has no leakage at spatial boundary, its integral on any Cauchy surface gives the symplectic form $\Omega$ of the theory:
\eqs{\Omega[\de_1,\de_2]=\int_\Sigma\boldsymbol{\omega}[\de_1,\de_2]}
The symplectic structure is conserved, i.e. it is s independent  of $\Sigma$ if the following quantity$-$the \emph{leakage}$-$vanishes at large $\rho$:
\eqs{\int_B\boldsymbol{\omega}[\de_1,\de_2]\label{leakage}}
where the integration is performed on a section of the asymptotic $dS_{d-1}$ space between $\tau_1$ and $\tau_2$. This is ensured if the asymptotic fall-off of the symplectic current is faster than $\omega^\rho\sim \rho^{1-d}$. In this case, one can associate conserved charges to gauge transformations of the theory.

Substituting \eqref{bnyc1} and \eqref{bnyc2} boundary conditions in \eqref{symp. current} and \eqref{leakage} reveals that the flux is vanishing only if $d>2p+4$, with odd outcome of excluding conserved charges for  electrodynamics in four dimensions. As it turns out, however, refining the symplectic current and boundary conditions will make the boundary leakage vanish for $d=2p+4$.

To discuss conservation of symplectic form we begin with decomposing the symplectic current \eqref{symp. current} as follows
\begin{align}
-(p+1)!\boldsymbol{\omega}^\mu&=\delta_1 \A_{\nu_0\cdots \nu_p}\delta_2\F^{\mu\nu_0\cdots \nu_p}-1\leftrightarrow 2\\
&=\delta_1 \A_{\nu_0\cdots \nu_p}\nabla^{\mu}\delta_2 \A^{\nu_0\cdots \nu_p}-
p\nabla^{[\nu_0}\left(
\delta_1\A_{\nu_0\cdots \nu_p}
\delta_2\A^{|\mu|\nu_1\cdots \nu_p]}
\right)\nonumber\\&\qquad\qquad+
p\left(\nabla^{[\nu_0}
\delta_1\A_{\nu_0\cdots \nu_p}\right)
\delta_2\A^{|\mu|\nu_1\cdots \nu_p]}
-1\leftrightarrow 2\label{omega decompose}\,.
\end{align}
Note that the second term in \eqref{omega decompose} is a total derivative. In specific conditions, if the total derivative term is pushed to the right-hand-side, the flux of the modified symplectic current is vanishing. To make this happen, fixing the Lorenz gauge
\eqs{
\nabla^{\nu_0}\A_{\nu_0\cdots\nu_p}=0
}
is sufficient to omit the third term in \eqref{omega decompose}. Regarding the first term, the $\rho$ component is
\eqs{
\de_1 \A_{a_0\cdots a_p}\nabla^\rho
\de_2 \A^{a_0\cdots a_p}-1\leftrightarrow 2\label{leakage term}
}
One can check that this term is $\ordr{2(2+p-d)}$. To see this, first note that the Christoffel symbols drop out by anti-symmetry in $\de_{1,2}$, if the number of upper indices inside the covariant derivative are the same for $\nabla\delta_1\A$ and $\nabla\delta_2\A$. Second, the partial derivative reduces the power by one unit, but the resulting term is symmetric in $\de_1\leftrightarrow\de_2$. Therefore the first non-vanishing term comes from $\rho$-derivative of subleading terms (if any), with overall power: $(2-d)\times2-2+2p+2=2(2+p-d)$. Consequently, taking $\rho^{d-1}$ factor from the metric, \eqref{leakage term} is irrelevant to the flux \eqref{leakage}, if $d>3+2p$. If this condition is satisfied, the conserved symplectic current augmented by the boundary term becomes
\begin{equation}
-(p+1)!\boldsymbol{\omega}^\mu=\de_1 \A_{\nu_0\cdots \nu_p}\delta_2\F^{\mu\nu_0\cdots \nu_p}+
(p+1)\nabla^{[\nu_0}\left(
\delta_1\A_{\nu_0\cdots \nu_p}
\delta_2\A^{|\mu|\nu_1\cdots \nu_p]}
\right)-1\leftrightarrow 2\label{final omega}
\end{equation}

\subsection{Residual gauge symmetry charges}
Given the symplectic form $\Omega$, the Hamiltonian of a variation $\de_{\boldsymbol{\Lambda}} \A$ induced by a gauge transformation $\A\to \A+\di\boldsymbol{\Lambda}$ is given by
\eqs{
\de Q_{\Lambda}[\A]=\Omega[\de_{\boldsymbol{\Lambda}},\de,\A]=\int_\Sigma\boldsymbol{ \omega}[\de_{\boldsymbol{\Lambda}} ,\de, \A]\,.
}
where $\A$ is the background solution at which the charge variation has been calculated. The symplectic current then is a total derivative $\omega^\mu=\nabla_\nu k^{\nu\mu}$ on-shell.
\begin{equation}
p!k_{\blaa}^{\mu\nu}\approx\frac{1}{2}\boldsymbol{\Laa}_{\alpha_1\cdots \alpha_{p}}\de \F^{\mu\nu\alpha_1\cdots \alpha_p}-(\di\boldsymbol{\Laa})^{\mu\alpha_1\cdots \alpha_p} {\de \A^{\nu}}_{\alpha_1\cdots \alpha_p}
-\mu\leftrightarrow \nu\label{charge density}\,.
\end{equation}
(We use index $\blaa$ for the charge, since only the leading term contributes.) The charge must be integrated on the boundary of $\Sigma$, a $(d-2)$-surface
\eqs{
\de Q_{\blaa}[\A]=\int_{C}\sqrt{\mathscr{G}}\,k^{\rho\tau}_{\blaa}\label{charge}\,.
}
The charge variation, in terms of de Sitter forms, is hence
\eqs{
\de Q_{\blaa}[\A]=\int_{C}\sqrt{\mathscr{G}}\,\Bigg[\boldsymbol{\laa}\wedge\star\de\mathbf{F}_\rho+\di\boldsymbol{\laa}_\rho\wedge\star\de\mathbf{A}+
\de\mathbf{A}_{\rho}\wedge\star(\di\boldsymbol{\laa})
\Big]\,,
}
where $\laa$ is defined in \eqref{bpgt3}.
An integration by parts in the second term gives
\eqs{
\di\boldsymbol{\laa}_\rho\wedge\star\delta\mathbf{A}=\di\Big(\boldsymbol{\laa}_\rho\wedge\star\delta\mathbf{A}\Big)-(-1)^{p-1}\boldsymbol{\laa}_\rho\wedge\di\star\delta\mathbf{A}\,.
}
Inside the integral, the total derivative on de sitter will be pulled back to the sphere and will drop by Stokes theorem. Then, from Lorenz gauge \eqref{Lorentz gauge split} one has
\eqs{
\di\star \delta\mathbf{A}=2(-1)^{p+1}\delta\mathbf{A}_\rho\,,
}
which leads to the following formula for charge variation,
\eqs{
\de Q_{\blaa}[\A]=\int_{C}\sqrt{\mathscr{G}}\,\Big[\boldsymbol{\laa}\wedge\star\de\mathbf{F}_\rho-2\boldsymbol{\laa}_\rho\wedge\star\de\mathbf{A}_\rho+
\de\mathbf{A}_{\rho}\wedge\star\di\boldsymbol{\laa}
\Big]\,,
}
in terms of forms on sphere. Being linear in field variations, one can readily verify the integrability condition
\eqs{
(\de_1\de_2-\de_2\de_1)Q_{\blaa}[\A]=0.
}
The charge variation can be hence be integrated to give a function $Q_{\blaa}$ on phase space, which after writing in terms of forms on sphere recovers \eqref{charge-expression-form} for the charge.

 

\providecommand{\href}[2]{#2}\begingroup\raggedright\endgroup
\end{document}